\documentstyle[11pt,aasms4]{article}

\newcommand{\qrmsps}{Q_{\rm rms-PS}}
\newcommand{\gap}{\;_\sim^>\;}
\newcommand{\lap}{\;_\sim^<\;}

\begin{document}

\vspace{-2.0truecm}
\begin{flushright}
KSUPT-97/1, KUNS-1448 \hskip 0.5cm May 1997
\end{flushright}
\vspace{-0.5truecm}

\title{MAX 4 and MAX 5 CMB anisotropy measurement constraints on
  open and flat-$\Lambda$ CDM cosmogonies}
\author{
  Ken Ganga\altaffilmark{1}, 
  Bharat Ratra\altaffilmark{2}, 
  Mark A. Lim\altaffilmark{3},
  Naoshi Sugiyama\altaffilmark{4}, and
  Stacy T. Tanaka\altaffilmark{5}
  }
\altaffiltext{1}{IPAC, MS 100--22, California Institute of Technology, 
                 Pasadena, CA 91125.}
\altaffiltext{2}{Department of Physics, Kansas State University, 
                 Manhattan, KS 66506.}
\altaffiltext{3}{Department of Physics, University of California, 
                 Santa Barbara, CA 93106.}
\altaffiltext{4}{Department of Physics, Kyoto University,
                 Kitashirakawa-Oiwakecho, Sakyo-ku, Kyoto 606.}
\altaffiltext{5}{Department of Physics, University of California,
                 Berkeley, CA 94720.}

\begin{abstract}
  We account for experimental and observational uncertainties in
  likelihood analyses of cosmic microwave background (CMB) anisotropy
  data from the MAX 4 and MAX 5 experiments. These analyses use CMB
  anisotropy spectra predicted in open and
  spatially-flat $\Lambda$ cold dark matter cosmogonies. Amongst the
  models considered, the combined MAX data set is most consistent with
  the CMB anisotropy shape in $\Omega_0 \sim 0.1-0.2$ open models and
  less so with that in old ($t_0 \gap$ 15 $-$ 16 Gyr, i.e., low $h$),
  high baryon density ($\Omega_B \gap$ 0.0175$h^{-2}$), low density
  ($\Omega_0 \sim$ 0.2 $-$ 0.4), flat-$\Lambda$ models. The MAX data
  alone do not rule out any of the models we consider at the $2\sigma$
  level. 
  
  Model normalizations deduced from the combined MAX data are consistent with 
  those drawn from the UCSB South Pole 1994 data, except for the flat
  bandpower model where MAX favours a higher
  normalization. The combined MAX data normalization for open models
  with $\Omega_0 \sim 0.1-0.2$ is higher than the upper $2\sigma$
  value of the DMR normalization. The combined MAX data normalization
  for old (low $h$), high baryon density, low-density flat-$\Lambda$
  models is below the lower $2\sigma$ value of the DMR normalization.
  Open models with $\Omega_0 \sim 0.4-0.5$ are not far from the shape most
  favoured by the MAX data, and for these models the MAX and DMR
  normalizations overlap. The MAX and DMR normalizations also overlap 
  for $\Omega_0 = 1$ and some higher $h$, lower $\Omega_B$, low-density
  flat-$\Lambda$ models. 
\end{abstract}


\section{Introduction}

Recent measurements of spatial anisotropy in the cosmic microwave
background indicate that CMB anisotropy data will soon provide useful
constraints on cosmological parameters such as $\Omega_0$, $h$, and
$\Omega_B$ (Bennett et al. 1996; Ganga et al. 1994; Guti\'errez et al.
1997; Piccirillo et al. 1997; Netterfield et al. 1997; Gundersen et
al. 1995; Tucker at al. 1997; Platt et al. 1997; Masi et al. 1996;
Lim et al. 1996, hereafter L96; Cheng et al.  1997; Griffin et al.
1997; Scott et al. 1996; Leitch et al. 1997; Church et al. 1997, see
Page 1997 for a review). Here, the present value of the (total)
nonrelativistic-mass density parameter $\Omega_0$, is $8\pi G\rho_b
(t_0)/(3H_0{}^2)$, where $G$ is the gravitational constant,
$\rho_b(t_0)$ is the mean nonrelativistic-mass density now, $H_0 =100 h$ km
s$^{-1}$ Mpc$^{-1}$ is the Hubble parameter now, and $\Omega_B$ is the
current value of the baryonic-mass density parameter.

Ganga et al. (1997a, hereafter GRGS) developed general methods to
account for experimental and observational uncertainties, such as
beamwidth and calibration uncertainties, in likelihood analysis of CMB
data sets.  These methods have previously been used in conjunction
with theoretically-predicted CMB anisotropy spectra in analyses of the
Gundersen et al. (1995) SP94 data and the Church et al. (1997) SuZIE
data (GRGS; Ganga et al. 1997b). Bond \&\ Jaffe (1997) have also
analyzed the SP94 data and the Netterfield et al. (1997) SK data.

In this paper we present results from a similar analysis of the MAX 4
and MAX 5 CMB anisotropy data sets (Devlin et al. 1994; Clapp et al.
1994; Tanaka et al. 1996, hereafter T96; L96).
MAX 4 and MAX 5 are the most recent of the published balloon-borne MAX
CMB anisotropy experiments. The original MAX detector and the MAX 1
results are discussed in Fischer et al. (1992). The ACME
telescope used in these MAX experiments is described in Meinhold
et al.  (1993a). The MAX 2 observational results are in Alsop et al.
(1992). Meinhold et al. (1993b) and Gundersen et al. (1993) present
the MAX 3 results. Analyses of the MAX 3 observations, in the context
of the fiducial CDM model, is given in Srednicki et al. (1993) and
Dodelson \& Stebbins (1994).

Descriptions of MAX 4 and 5 are given in Devlin et al. (1994), Clapp
et al.  (1994), T96, and L96; we review here the information needed
for our analyses.  Data were taken in four frequency bands centered at
3.5, 6, 9, and 14 cm$^{-1}$; our analyses do not use the 14 cm$^{-1}$ 
band data.  The FWHM of the beams, assumed to be gaussian, are:
$0.55^\circ \pm 0.05^\circ$ for the MAX 4 3.5 cm$^{-1}$ band, $0.75^\circ \pm
0.05^\circ$ for the MAX 4 6/9 cm$^{-1}$ bands, $0.5 (1 \pm 0.10)^\circ$ for the 
MAX 5 3.5 cm$^{-1}$ band, and $0.55 (1 \pm 0.10)^\circ$ for the 
MAX 5 6/9 cm$^{-1}$ bands, where the uncertainties are one standard deviation. 
The MAX data used here were taken during smooth, constant velocity, constant
declination, azimuthal scans extending $6^\circ$ on the sky for MAX 4 
and $8^\circ$ on the sky for MAX 5. While observing, the beam was
sinusoidally chopped with a half peak-to-peak amplitude of $1.4^\circ$
on the sky. The data were coadded into 21 bins for MAX 4 and 29 bins for MAX 5.

MAX 4 data were taken in smooth scans of $\pm 3^\circ$ on the sky
centered near the stars $\gamma$ Ursae Minoris, $\iota$ Draconis
(hereafter 4ID), and $\sigma$ Herculis (hereafter 4SH). Sky rotation
has a significant effect on the $\gamma$ Ursae Minoris scan pattern
(Devlin et al. 1994), so we do not analyze this data set here. The bin
positions in the released 4ID and 4SH data sets are data-weighted
averages (A. Clapp, private communication 1996); we use evenly spaced bins in 
this analysis. MAX 5 data were taken in smooth scans of $\pm 4^\circ$
on the sky centered near the stars HR5127 (hereafter 5HR), $\mu$
Pegasi (hereafter 5MP), and $\phi$ Herculis (hereafter 5PH). The 9
cm$^{-1}$ 5PH data is thought to contain atmospheric emission (T96,
5PH data was taken at lower balloon altitude), and is not used for CMB
anisotropy analyses.  The 5MP data has structure that correlates with
$IRAS$ 100 $\mu$m dust emission (L96). Off-diagonal noise correlations
affect the 5HR and 5PH results by $\lap 2\%$ (T96), and are ignored
here.

MAX 4 and 5 were calibrated primarily by using a membrane transfer
standard (Fischer et al. 1992); the absolute calibration uncertainty is
10\% ($1\sigma$).

In $\S$\ref{sec:computation} we summarize some of the computational
techniques used in our analysis. Our results and a discussion are in
$\S$\ref{sec:results}, and we conclude in $\S$\ref{sec:conclusion}.

\section{Summary of Computation}\label{sec:computation}

Our conventions, notation, and techniques are those of GRGS.

The CMB anisotropy spectra for some of the models we consider
here are shown in Figure~\ref{fig:spectra}. The models are described
in Ratra et al.  (1997) and GRGS, and the computation of the spectra
is discussed in Sugiyama (1995). Besides the flat bandpower and
fiducial CDM models, we also consider
open and flat-$\Lambda$ CDM cosmogonies. The
models assume gaussian, adiabatic, primordial energy-density power
spectra. The flat-$\Lambda$ models assume a scale-invariant
energy-density power spectrum (Harrison 1970; Peebles \&\ Yu 1970;
Zel'dovich 1972), as is found in the simplest spatially-flat inflation
models (Guth 1981, also see Kazanas 1980; Sato 1981a,b). The
open models assume the energy-density power spectrum (Ratra \& Peebles
1994, 1995; Bucher, Goldhaber, \& Turok 1995; Yamamoto, Sasaki, \&
Tanaka 1995) of the simplest open-bubble inflation models (Gott 1982;
Guth \& Weinberg 1983). The model spectra are parameterized by the
quadrupole-moment amplitude of the CMB anisotropy, $Q_{\rm rms-PS}$,
as well as $\Omega_0$, $h$, and $\Omega_B$. The spectra shown in
Figure~\ref{fig:spectra} are normalized to the DMR maps (G\'orski et al. 
1996,1998; Stompor 1997). Parameter values for the models considered are 
given in Table~6.

The low-density models considered here are the simplest ones roughly
consistent with most current observations. For flat-$\Lambda$ models see
Kitayama \& Suto (1996), Ratra et al. (1997), Bunn \& White (1997), Turner 
(1997), Peacock (1997), and Cole et al. (1997), and for the open case see
Kitayama \& Suto (1996), Ratra et al. (1997), G\'orski et al (1998),
Gott (1997), Peacock (1997), and Cole et al. (1997).
The values of the parameters $\Omega_0$, $h$, and $\Omega_B$ used here
are chosen to be roughly consistent with present observational
estimates of $\Omega_0$, $h$, the age of the universe, and the
constraints on $\Omega_B$ that follow from standard nucleosynthesis
(Ratra et al. 1997). In this analysis we ignore the effects of tilt,
primordial gravity waves, and reionization.  These effects are
unlikely to be significant in viable open models.\footnote{
Maia \& Lima (1996), Tanaka \& Sasaki (1997), and Bucher \& Cohn (1997)
have studied primordial gravity waves in the open model. Initial indications 
are that predictions based on the simplest observationally-viable 
single-scalar-field open-bubble-inflation energy-density power spectrum (Ratra 
\& Peebles 1994, 1995) are negligibly affected by primordial gravity waves.}
In order to
reconcile some of the flat-$\Lambda$ models with observational data, 
however, some such effect is needed to suppress intermediate-scale (CMB and 
matter) and small-scale (matter) power (Stompor,
G\'orski, \& Banday 1995; Ostriker \&\ Steinhardt 1995; Ratra et al.
1997; Klypin, Primack, \&\ Holtzman 1996; Ganga, Ratra, \& Sugiyama
1996; Maddox, Efstathiou, \& Sutherland 1996; Peacock 1997; Cole et
al. 1997).

Figure~1 also shows the four different zero-lag window functions,
$W_l$, for the individual MAX channels at their nominal beamwidths. The
usual window function parameters, the value of $l$ where $W_l$ is
largest, $l_{\rm m}$, the two values of $l$ where $W_{l_{e^{-0.5}}} =
e^{-0.5} W_{l_{\rm m}}$, $l_{e^{-0.5}}$, and the effective multipole,
$l_{\rm e}$, are given in Table~1 for these window functions (e.g.,
Bond 1996; GRGS).

Figure~2 shows the moments $(\delta T_{\rm rms}{}^2)_l$ for two MAX
window functions and some selected CMB anisotropy spectra (GRGS, eqs.
[5] \& [6]). Given an assumed CMB anisotropy spectrum, these moments
provide a convenient measure of the range of $l$ to which MAX is
sensitive. Table~2 gives $l_{\rm m}$, the value of $l$ at which
$(\delta T_{\rm rms}{}^2)_l$ is largest, and $l_{e^{-0.5}}$ the two
multipoles where $(\delta T_{\rm rms} {}^2)_{l_{e^{-0.5}}} =
e^{-0.5}(\delta T_{\rm rms}{}^2)_{l_{\rm m}}$, for the MAX window
functions and the models we consider. As is true for SP94 and SuZIE,
the range of multipole moments to which each window function is
sensitive is quite model dependent (GRGS; Ganga et al. 1997b).

The reduced MAX 4 and MAX 5 data are shown in Figure 3. In Table~3,
the column labelled ``Sky'' indicates the estimated anisotropy rms for
those individual channel data sets thought to be purely CMB
anisotropy. This is computed from the data of Figure 3 as the square
root of the difference between the variance of the mean temperatures
and the variance of the error bars.

The computation of the likelihood function is described in GRGS. The
only difference between what is done here and in GRGS is that for the
5MP data (but not the other data sets) we also marginalized over a
possible dust contamination given by the spectrum in L96 and an
arbitrary spatial morphology. Beamwidth and calibration uncertainties are 
accounted for as described in GRGS.  We again use three-point Gauss-Hermite 
quadrature to marginalize over beamwidth uncertainty.

In Table 3, the column labelled ``FBP'' lists central $\delta T_{\rm
  rms}$ values derived from likelihood analyses using the flat
bandpower (FBP) spectrum.

To derive the $Q_{\rm rms-PS}$ central value and limits from the
likelihood function we assume a uniform prior in $Q_{\rm rms-PS} (\geq
0)$.  The central value is taken to be the value of $\qrmsps$ at which
the probability density distribution peaks. The MAX limits we quote are $\pm
1\sigma$ highest posterior density (HPD) limits for ($2\sigma$ HPD)
detections.  For nondetections we quote upper $2\sigma$ equal tail
(ET) limits.\footnote{
ET limits are determined by integrating the probability density distribution
function starting from 0 $\mu$K. The $2\sigma$ ET limit encompasses $97.7\%$
of the area. HPD limits are determined by integrating the probability density
distribution function starting at the peak and minimizing the difference
between the upper and lower limits. The $1\sigma$ and $2\sigma$ HPD limits
encompass $68.3\%$ and $95.5\%$ of the area, respectively. See GRGS for 
a discussion.}

In Table 4 we give bandtemperature ($\delta T_l = \delta T_{\rm rms}/
\left[\sum^\infty_{l=2} \left[(l + 0.5) W_l /\{l (l+1)\}\right]\right]^{0.5}$,
e.g., Bond 1996; GRGS, eq. 
[7]) central values and $\pm 1\sigma$ limits derived from flat bandpower 
likelihood analyses of the individual-channel MAX data sets. The last
two columns give the average of the $\pm 1\sigma$ error bars in $\mu$K
and as a percentage of the central value. In Table 5 we give the
corresponding numerical values derived from flat bandpower likelihood
analyses of the combined-channel MAX data sets.

Central values and limits for $\qrmsps$ are given in Tables 6 and 7
for the various individual and combined MAX data sets. Some of the likelihood
functions used to derive these numerical values are shown in Figure 4.
Tables 6 and 7 also show the results from analyses of the DMR data,
accounting for both statistical and systematic DMR uncertainties
(G\'orski et al.  1996,1998; Stompor 1997). The DMR limits quoted are 
$\pm 2\sigma$ HPD for the two extreme data sets:
(1) galactic-frame maps accounting for the high-latitude Galactic emission
correction and including the $l = 2$ moment in the analysis; and (2)
ecliptic-frame maps ignoring the high-latitude Galactic correction and
excluding the $l = 2$ moment. The DMR central values quoted are the 
arithmetic mean of the $\pm 2\sigma$ limits. See G\'orski et al. (1998) 
for a discussion of the DMR central values and limits.

Tables 8 and 9 give the values of the probability density distribution
functions at the peak, and the marginalized (over $Q_{\rm rms-PS}$)
probability density distribution values for the various
combined-channel MAX data sets.\footnote{
These values are computed using a uniform prior in $Q_{\rm rms-PS}$.
The following conclusions are therefore based solely on the MAX data.}
The models shown in the figures were
chosen on the basis of their marginal probability distribution values
for the C(ombined) MAX data set.  Model O1 (an $\Omega_0 = 0.1$, open
model) is the most likely model, followed, among the selected models,
by Flat (flat bandpower), O14 (fiducial CDM), models $\Lambda$2 and
O11 (a spatially-flat, $\Omega_0 = 0.2$ model and an open, $\Omega_0 =
0.5$ model), and the least likely one, $\Lambda$10 (a flat-$\Lambda$,
$\Omega_0 = 0.4$ model). These selected models include the most and
least likely open and flat-$\Lambda$ ones among those considered.

\section{Results and Discussion}\label{sec:results}

From Table 3 we see that, for those channels with $2\sigma$ HPD detections, the 
rms estimated from likelihood analyses using flat bandpower spectra is in good 
agreement with the ``sky'' rms estimated from the data, especially for MAX 5.
The exception is 4ID 6 cm$^{-1}$; based on the SP94 analyses (GRGS),
it is unlikely that this can significantly affect conclusions drawn
from the combined MAX data sets.

Tables 4 and 5 give the central values and $\pm 1\sigma$ limits on
bandtemperature derived from likelihood analyses with the flat
bandpower spectrum. These numerical values can not be compared
directly to those of T96 and L96 because of the many differences in
the analyses.  For instance, beamwidth and calibration uncertainties
are accounted for in different ways, and T96 use a likelihood ratio
method while we use a maximum likelihood method (see T96 for
discussion).  Nevertheless, it is reassuring that the largest
differences are not greater than $\sim 1\sigma$.

From Tables 4 and 5, we note that the deduced 4SH average absolute
uncertainty is larger than that of the 4SH 9 cm$^{-1}$ channel and only
slightly smaller than the 4SH 3.5 cm$^{-1}$ one. Similarly, the
deduced 5PH average absolute uncertainty is only slightly smaller than
the 5PH 3.5 and 6 cm$^{-1}$ ones. On the other hand, the average
absolute uncertainties do shrink when the 4ID or 5HR
individual-channel data sets are combined.
 
These error bars only account for instrumental and atmospheric noise, sample 
variance due to the limited number of independent data pixels, and beamwidth 
and calibration uncertainty. For purely CMB anisotropy data, instrumental
and atmospheric noise should integrate down with more
channels of data, sample variance should not, and the beamwidth and
calibration uncertainties are negligible for the purposes of this
discussion. Since the MAX 4 3.5 and 6/9 cm$^{-1}$ window functions are
rather dissimilar, an analysis of the behaviour of the error bars in
this case will require a numerical simulation. We focus here on the
5HR and 5PH data sets.

We follow the analysis of \S 3 of GRGS. In their notation, the total
average absolute $\delta T_l$ error bars for the combined-channel
($\sigma_{\rm tot., com.}$) and individual-channel ($\sigma_{\rm tot.,
  ind.}$) data sets are, from Tables 4 and 5,
\begin{eqnarray}
  \sigma^{\rm 5HR}_{\rm tot., ind.} = & 13\ \mu{\rm K}, \qquad
    \sigma^{\rm 5HR}_{\rm tot., com.} & = 8.4\ \mu{\rm K}, \nonumber \\
  \sigma^{\rm 5PH}_{\rm tot., ind.} = & 22\ \mu{\rm K}, \qquad
    \sigma^{\rm 5PH}_{\rm tot., com.} & = 21\  \mu{\rm K}.
\end{eqnarray}
To get this behaviour, the sample variance ($\sigma_{\rm SV}$) and
intrinsic noise ($\sigma_{\rm N}$) contributions to the $\delta T_l$
error bars need to be
\begin{eqnarray}
  \sigma^{\rm 5HR}_{\rm SV} = & 5.0\ \mu{\rm K}, \qquad
          \sigma^{\rm 5HR}_{\rm N}  & = 12\ \mu{\rm K}, \nonumber \\
  \sigma^{\rm 5PH}_{\rm SV}  = & 20\ \mu{\rm K}, \qquad
          \sigma^{\rm 5PH}_{\rm N}   & =  9.3\ \mu{\rm K};
\end{eqnarray}
i.e., the 5HR data needs to be dominated by intrinsic noise while the
5PH data needs to be dominated by sample variance. To see if this is
reasonable, we now estimate the MAX 5 sample variance, following the
discussion of GRGS based on the simulations of Netterfield et al.
(1995). We approximate the MAX 5 individual-channel beamwidths by
$\sigma_{\rm FWHM} = 0.5^\circ$; the $8^\circ$ scans then have 16
independent pixels (with 29 bins, MAX 5 is quite oversampled). With
the Netterfield et al. (1995) 10\% simulation correction to the
analytic estimate of sample variance (GRGS), the MAX 5 sample variance
is 19\% of the $\delta T_l$ central value. Using this, the 5HR and 5PH 
$\delta T_l$ central values of
Table 5, and the standard equations, we find for the sample variance
and intrinsic noise contributions to the 5HR and 5PH $\delta T_l$
average error bars,
\begin{eqnarray}
  \sigma^{\rm 5HR}_{\rm SV} = & 5.3\ \mu{\rm K}, \qquad
             \sigma^{\rm 5HR}_{\rm N}  & = 12\ \mu{\rm K},  \nonumber \\
  \sigma^{\rm 5PH}_{\rm SV}  = & 14\ \mu{\rm K}, \qquad
             \sigma^{\rm 5PH}_{\rm N}   & = 17\ \mu{\rm K}.
\end{eqnarray}
This approximate estimate of the 5HR sample variance and intrinsic
noise is very close to what is needed (eq. [2]) to explain the
behaviour of the 5HR error bars. However, eq. (3) indicates that the
5PH sample variance is not as large as needed (eq. [2]) to explain the
behaviour of the 5PH error bars. While it is premature to read
much from such an approximate analysis (GRGS), we note that the 5PH
data were taken at lower balloon altitude and that the 5PH 9 cm$^{-1}$
data can not be used for analyses of CMB anisotropy (T96).

We do not record here $\delta T_l$ numerical values for other models
from the individual-channel MAX data sets (the flat bandpower model
values are in Table 4). As was the case for SP94 (GRGS), for a given
MAX data set the deduced $\delta T_l$ values vary by $\sim 0$-10\%
from model to model, with a typical range of $\sim 5\%$. This is
significantly smaller than the variations for SuZIE (Ganga et al.
1997b).

The 5MP data do not show a $2\sigma$ HPD detection. Because of the
significant non-CMB component in this data set (L96), and the need for
``subtraction" prior to CMB anisotropy analysis (L96), there is the worry
that this data set might be biased. Fortunately, as is seen from the
5, 5noMP, C(ombined), and CnoMP entries in Table 5,
inclusion/exclusion of 5MP shifts the deduced normalization only by
$\sim 0.8\sigma$, so 5MP does not significantly affect the final
results. We note that the 5MP upper limit is quite consistent with the
4ID and 5HR detections, but is somewhat below the 4SH and 5PH
detections. Given the discussion about sample variance and intrinsic
noise uncertainties above, we believe that it is more reasonable to
include the 5MP results.

Tables 6 -- 9 summarize our MAX data set analyses.  

Table 8 lists the maximum values of the probability density
distribution function for the various combined-channel MAX data sets
and all models considered here. From these numerical values, and from
Figure 4, one sees that for all the combined-channel MAX data sets,
except 5MP which does not have a detection, the likelihood functions
are peaked and very well separated from 0 $\mu$K. Note the 14\% error
bar for the C(ombined) data set in Table 5; the combined MAX data
shows a very significant detection, even after calibration and
beamwidth uncertainties are accounted for. For comparison, depending
on model, the corresponding DMR error bars are $\sim 10-12\%$
(G\'orski et al. 1998). Given such likelihood functions, it is
reasonable to choose between models on the basis of the value of the
marginal probability distribution function.

Table 9 lists the marginal probability distribution values for the
combined-channel MAX data sets. In all cases, among all the models
considered here, the MAX data favour low-density open models with
$\Omega_0 \sim 0.1-0.2$ (models O1 -- O4).\footnote{
Figure 4a shows that the likelihood functions for all the combined-channel
MAX data sets are significantly nongaussian. Consequently, conclusions
about model viability based on the marginal probability density
need not be identical to those drawn from the projected probability 
density. From Tables 8 and 9 one sees that the projected probability 
density does not distinguish as much between models as does the 
marginal probability distribution, and in fact weakly favours 
$\Omega_0 = 1$ over the open $\Omega_0 = 0.1$ case for the C(ombined)
data.}
This is consistent with what we find from all the individual-channel MAX data 
sets (not shown here), and with what is found from an analysis of the SP94 data
(GRGS). Our marginal probability distribution function is evaluated at
isolated points in model-parameter ($\Omega_0$, $h$, $\Omega_B h^2$)
space; assuming it is a gaussian, a model $1\sigma$ away from the most
favoured low-density open model CMB anisotropy shape has a marginal
value of 0.61, and a model with marginal value 0.38 is $1.4\sigma$
away from the most favoured low-density open model. The C(ombined) MAX
data set values of Table 9 do distinguish between models, although not
at a very significant level.

We conclude that the MAX data are most consistent with the CMB
anisotropy shape in low-density open CDM models with $\Omega_0 \sim$
0.1 -- 0.3 and 0.4 -- 0.5 (with larger $h$ and smaller $\Omega_B$),
and with the flat bandpower shape, among the models we consider here.
The fiducial CDM and flat-$\Lambda$ models have CMB anisotropy
spectral shapes which MAX does not favour. This is especially true for
old ($t_0 \gap$ 15 -- 16 Gyr), large baryon density ($\Omega_B \gap
0.0175 h^{-2}$), low-density ($\Omega_0 \sim$ 0.2 -- 0.4),
flat-$\Lambda$ models. These results are consistent with those based
on an analysis of the SP94 data (GRGS), as well as with earlier
analyses of multiple CMB anisotropy data points (Ratra et al. 1997;
Ganga et al. 1996; also see Hancock et al. 1997). We emphasize that,
under the gaussian marginal assumption, the MAX data alone do not rule
out any of the models we consider here at the $2\sigma$ level.

Tables 6 and 7 list the $Q_{\rm rms-PS}$ values derived from the
various combined-channel MAX data sets. These normalizations are in
striking agreement with those deduced from the SP94 data (GRGS),
except for the flat bandpower model where MAX favours a
higher normalization than does SP94. This is further confirmation
that the observed CMB anisotropy spectrum rises towards large
$l$ (Scott, Silk, \& White 1995; Ratra et al. 1997; Ganga et al. 1996; 
Netterfield et al. 1997; Hancock et al. 1997; Page 1997; Lineweaver et al. 1997; Rocha \& Hancock 1997).\footnote{
  The steepness of the rise towards large $l$ depends on which data 
  sets are included in the analysis. Hancock et al. (1997), Lineweaver et 
  al. (1997), and Rocha \& Hancock (1997) favour a slightly steeper $C_l$ 
  than do Ganga et al. (1996). This is because Ganga et al. (1996) include 
  a number of 
  data points (e.g., the four MSAM points and the MAX 3MP point which 
  is consistent with the repeat 5MP result and the 4ID and 
  5HR measurements) which favour shallower $C_l$ spectra.}

For open models with $\Omega_0 \sim 0.1-0.2$ the combined MAX data
normalization is above the upper $2\sigma$ DMR normalization (G\'orski
et al. 1998); for $\Omega_0 \sim 0.1$ it is also above the SuZIE
$2\sigma$ upper limit (Ganga et al. 1997b). For the older (smaller
$h$), higher $\Omega_B$, flat-$\Lambda$ models, the combined MAX
normalization is below the lower $2\sigma$ DMR normalization (Stompor
1997). Both of these results are consistent with earlier CMB anisotropy 
conclusions (Ratra et al. 1997; Ganga et al. 1996). An open model
with $\Omega_0 \sim 0.4-0.5$ and $h \sim 0.65$ has a CMB anisotropy
spectral shape which is not far from what is favoured by the MAX data;
in addition, the MAX normalization for such a model is consistent with
what is deduced from the DMR (G\'orski et al.  1998) and SP94 data
(GRGS), and is also consistent with the SuZIE $2\sigma$ upper limit
(Ganga et al. 1997b). The MAX and DMR normalizations also overlap for $\Omega_0
=1$ and some higher $h$, lower $\Omega_B$, low-density flat-$\Lambda$ 
models.

\section{Conclusion}\label{sec:conclusion}

We have accounted for beamwidth- and calibration-uncertainty in
likelihood analyses of the MAX 4 and MAX 5 observational data that
make use of theoretical CMB spatial anisotropy spectra in
open and spatially-flat $\Lambda$ CDM
cosmogonies. In our analyses, we have assumed that the appropriately
reduced MAX data is purely CMB spatial anisotropy. Other general
caveats may be found in \S 4 of GRGS, and it is prudent to bear these
in mind when interpreting our results.

The marginal probability distribution function values indicate that
the MAX data favour low-density open models over older, high
$\Omega_B$, low-density, flat-$\Lambda$ models. However, no model
considered here is ruled out at the $2\sigma$ level by the MAX data
alone, at least in the gaussian marginal probability distribution
approximation. Combined with results from the analyses of the DMR,
SP94, and SuZIE data, we find that $\Omega_0 \sim 0.1-0.2$ open models have
CMB anisotropy spectra shallower than what is favoured by the
observations while low $h$, high $\Omega_B$ flat-$\Lambda$ models have
spectra steeper than what is favoured by the observations, in
agreement with earlier analyses.

It is interesting that the MAX and SP94 data sets lead to almost
identical conclusions for observationally-motivated low-density open
and flat-$\Lambda$ CDM models. If confirmed, this result might be of
some passing significance.

\acknowledgments

We acknowledge helpful discussions with A. Clapp, J. Cohn, J. Gundersen, 
S. Hanany, A. Lee, G. Rocha, and R. Stompor, as well as, especially, M. Devlin 
and L. Page.  BR acknowledges support from NSF grant EPS-9550487 with
matching support from the state of Kansas and from a K$^*$STAR First
award. This work was partially carried out at the Infrared Processing and
Analysis Center and the Jet Propulsion Laboratory of the California Institute
of Technology, under a contract with the National Aeronautics and Space 
Administration.

\clearpage

\begin{table}
\begin{center}
\caption{Numerical Values for the Zero-Lag Window Function Parameters}
\vspace{0.3truecm}
\label{tab:winpar}
\begin{tabular}{cccccc}
\tableline\tableline
Channel             & $l_{e^{-0.5}}$ & $l_{\rm e}$ & $l_{\rm m}$  
                    & $l_{e^{-0.5}}$ & $\sqrt{I(W_l)}$  \\
\tableline
MAX 4 3.5 cm$^{-1}$ & 80 & 133 & 145 & 224 & 1.51 \\
MAX 4 6/9 cm$^{-1}$ & 70 & 114 & 127 & 196 & 1.41 \\
MAX 5 3.5 cm$^{-1}$ & 83 & 139 & 150 & 232 & 1.55 \\
MAX 5 6/9 cm$^{-1}$ & 80 & 133 & 146 & 224 & 1.51 \\
\tableline
\end{tabular}
\end{center}
\end{table}

\begin{table}
\renewcommand{\arraystretch}{0.75}
\begin{center}
\caption{Numerical Values for Parameters Characterizing the Shape of
$(\delta T_{\rm rms}{}^2)_l$}
\vspace{0.3truecm}
\label{tab:winpartwo}
\tablenotetext{{\rm a}}{$l_{e^{-0.5}}$ for the window.} 
\tablenotetext{{\rm b}}{$l_{\rm e}$ for the window.}
\begin{tabular}{ccrclrclrclrcl}
\tableline\tableline
$W_l$: &  & 
\multicolumn{3}{c}{MAX 4 3.5 cm$^{-1}$} &  
\multicolumn{3}{c}{MAX 4 6/9 cm$^{-1}$} & 
\multicolumn{3}{c}{MAX 5 3.5 cm$^{-1}$} & 
\multicolumn{3}{c}{MAX 5 6/9 cm$^{-1}$} \\
\tableline
$\#$ & $(\Omega_0 , h , \Omega_B h^2)$ & $l_{e^{-.5}}$ & $l_{\rm m}$ 
& $l_{e^{-.5}}$ & $l_{e^{-.5}}$ & $l_{\rm m}$ & $l_{e^{-.5}}$ 
& $l_{e^{-.5}}$ & $l_{\rm m}$ & $l_{e^{-.5}}$ & $l_{e^{-.5}}$
& $l_{\rm m}$ & $l_{e^{-.5}}$ \\
(1) & (2) & (3) & (4) & (5) & (6) & (7) & (8) & (9) & (10) & (11) & (12) & (13)
& (14) \\
\tableline
         O1 &(0.1, 0.75, 0.0125)& 54 & 130 & 213 & 42 & 110 & 184 & 57 & 135 & 221 & 54 & 130 & 214 \\
         O2 &(0.2, 0.65, 0.0175)& 66 & 139 & 222 & 54 & 119 & 192 & 69 & 144 & 230 & 66 & 139 & 222 \\
         O3 &(0.2, 0.70, 0.0125)& 62 & 135 & 219 & 50 & 115 & 189 & 65 & 141 & 228 & 62 & 135 & 220 \\
         O4 &(0.2, 0.75, 0.0075)& 57 & 131 & 216 & 47 & 111 & 185 & 60 & 136 & 224 & 57 & 131 & 216 \\
         O5 &(0.3, 0.60, 0.0175)& 70 & 143 & 226 & 58 & 122 & 195 & 73 & 149 & 235 & 70 & 143 & 227 \\
         O6 &(0.3, 0.65, 0.0125)& 65 & 139 & 224 & 54 & 118 & 193 & 69 & 145 & 233 & 66 & 140 & 224 \\
         O7 &(0.3, 0.70, 0.0075)& 60 & 135 & 222 & 50 & 114 & 189 & 63 & 140 & 230 & 61 & 135 & 222 \\
         O8 &(0.4, 0.60, 0.0175)& 71 & 146 & 229 & 59 & 124 & 198 & 75 & 152 & 237 & 71 & 146 & 229 \\
         O9 &(0.4, 0.65, 0.0125)& 67 & 143 & 227 & 55 & 120 & 195 & 70 & 149 & 235 & 67 & 143 & 227 \\
        O10 &(0.4, 0.70, 0.0075)& 61 & 139 & 226 & 50 & 115 & 193 & 64 & 145 & 234 & 61 & 139 & 226 \\
        O11 &(0.5, 0.55, 0.0175)& 75 & 149 & 230 & 62 & 127 & 200 & 78 & 155 & 237 & 75 & 150 & 230 \\
        O12 &(0.5, 0.60, 0.0125)& 70 & 147 & 229 & 57 & 124 & 198 & 74 & 153 & 236 & 70 & 148 & 229 \\
        O13 &(0.5, 0.65, 0.0075)& 64 & 145 & 228 & 52 & 120 & 197 & 68 & 151 & 236 & 65 & 145 & 228 \\
        O14 &(1.0, 0.50, 0.0125)& 81 & 150 & 216 & 66 & 133 & 196 & 85 & 154 & 220 & 81 & 151 & 216 \\ 
 $\Lambda$1 &(0.1, 0.90, 0.0125)& 85 & 149 & 214 & 74 & 133 & 194 & 88 & 152 & 219 & 85 & 149 & 214 \\
 $\Lambda$2 &(0.2, 0.80, 0.0075)& 85 & 149 & 213 & 73 & 134 & 194 & 88 & 153 & 218 & 85 & 150 & 213 \\
 $\Lambda$3 &(0.2, 0.75, 0.0125)& 85 & 150 & 215 & 74 & 133 & 195 & 88 & 153 & 220 & 86 & 150 & 215 \\
 $\Lambda$4 &(0.2, 0.70, 0.0175)& 86 & 151 & 217 & 75 & 134 & 196 & 89 & 155 & 222 & 86 & 151 & 217 \\
 $\Lambda$5 &(0.3, 0.70, 0.0075)& 85 & 150 & 215 & 71 & 134 & 195 & 88 & 154 & 219 & 85 & 151 & 215 \\
 $\Lambda$6 &(0.3, 0.65, 0.0125)& 85 & 151 & 216 & 73 & 134 & 196 & 88 & 154 & 221 & 85 & 151 & 216 \\
 $\Lambda$7 &(0.3, 0.60, 0.0175)& 86 & 151 & 219 & 74 & 134 & 197 & 89 & 156 & 224 & 86 & 151 & 219 \\
 $\Lambda$8 &(0.4, 0.65, 0.0075)& 84 & 151 & 215 & 70 & 134 & 195 & 87 & 155 & 220 & 84 & 151 & 215 \\
 $\Lambda$9 &(0.4, 0.60, 0.0125)& 85 & 151 & 216 & 72 & 134 & 196 & 88 & 155 & 221 & 85 & 151 & 217 \\
$\Lambda$10 &(0.4, 0.55, 0.0175)& 86 & 152 & 219 & 74 & 134 & 197 & 89 & 156 & 224 & 86 & 152 & 219 \\
$\Lambda$11 &(0.5, 0.60, 0.0125)& 84 & 151 & 215 & 71 & 134 & 195 & 87 & 155 & 220 & 84 & 151 & 215 \\
  Flat      &         ...       & 41 & 103 & 183 & 36 &  90 & 160 & 42 & 106 & 189 & 41 & 103 & 183 \\
\tableline
  $W_l$     &         ...       & 80\tablenotemark{a} & 133\tablenotemark{b} & 224\tablenotemark{a} & 70\tablenotemark{a} & 114\tablenotemark{b} & 
196\tablenotemark{a} & 83\tablenotemark{a} & 139\tablenotemark{b} & 
232\tablenotemark{a} & 80\tablenotemark{a} & 133\tablenotemark{b} & 
224\tablenotemark{a} \\
\tableline
\end{tabular}
\end{center}
\end{table}

\begin{table}
\begin{center}
\caption{Numerical Values for Rms Temperature Anisotropies\tablenotemark{a}}
\label{tab:rmsani}
\vspace{0.3truecm}
\tablenotetext{{\rm a}}{$\delta T_{\rm rms}$ in $\mu$K.}
\tablenotetext{{\rm b}}{Estimated from the data of Fig. 3, as discussed in 
 $\S 2$.}
\tablenotetext{{\rm c}}{Converted to rms from the results of 
 the likelihood analysis for the flat bandpower (FBP) angular spectrum,
 accounting for beamwidth and calibration uncertainties. Note that the 4ID 
 9 cm$^{-1}$ and 4SH 6 cm$^{-1}$ data sets do not have $2\sigma$ HPD 
 detections.}
\begin{tabular}{lcc}
\tableline\tableline
Channel & ``Sky"\tablenotemark{b} & FBP\tablenotemark{c} \\
\tableline
4ID 3.5 cm$^{-1}$ & 71 &  86 \\
4ID   6 cm$^{-1}$ & 29 &  95 \\
4ID   9 cm$^{-1}$ & 55 &  78 \\
4SH 3.5 cm$^{-1}$ & 89 & 130 \\
4SH   6 cm$^{-1}$ & 16 &   0 \\
4SH   9 cm$^{-1}$ & 69 &  75 \\
5HR 3.5 cm$^{-1}$ & 69 &  62 \\
5HR   6 cm$^{-1}$ & 44 &  40 \\
5HR   9 cm$^{-1}$ & 49 &  56 \\
5PH 3.5 cm$^{-1}$ & 64 &  78 \\
5PH   6 cm$^{-1}$ & 85 & 110 \\
\tableline
\end{tabular}
\end{center}
\end{table}

\begin{table}
\renewcommand{\arraystretch}{0.8}
\begin{center}
\caption{Numerical Values for Bandtemperature\tablenotemark{a}{ } 
from Likelihood Analyses Assuming a Flat Bandpower Spectrum, for the 
Individual-Channel Data Sets}
\vspace{0.3truecm}
\label{tab:bandtemp}
\tablenotetext{{\rm a}}{$\delta T_{l}$ in $\mu$K.}
\tablenotetext{{\rm b}}{Average absolute error in $\mu$K.}
\tablenotetext{{\rm c}}{Average fractional error, as a fraction of the 
central value.}
\tablenotetext{{\rm d}}{4ID 9 cm$^{-1}$ and 4SH 6 cm$^{-1}$ do not have 
$2\sigma$ HPD detections; the appropriate ET $2\sigma$ upper limits are
220 $\mu$K and 170 $\mu$K respectively.}
\tablenotetext{{\rm e}}{Data from 5HR and 5PH.}
\begin{tabular}{lccccc}
\tableline\tableline
Data Set & $-1\sigma$ & Peak & $+1\sigma$ &
Ave. Abs. Err.\tablenotemark{b} & Ave. Fra. Err.\tablenotemark{c} \\
\tableline
4ID   3.5 cm$^{-1}$                  & 37 & 57 &  86 & $\pm 25$  & $\pm 43$\% \\
4ID     6 cm$^{-1}$                  & 38 & 67 & 110 & $\pm 37$  & $\pm 55$\% \\
4ID 9 cm$^{-1}$ \tablenotemark{d}    & 25 & 55 & 110 & $\pm 40$  & $\pm 73$\% \\
4SH   3.5 cm$^{-1}$                  & 58 & 86 & 130 & $\pm 35$  & $\pm 41$\% \\
4SH 6 cm$^{-1}$ \tablenotemark{d}    &  0 &  0 &  63 & $\pm 32$  & ...        \\
4SH     9 cm$^{-1}$                  & 30 & 53 &  91 & $\pm 31$  & $\pm 58$\% \\
5HR   3.5 cm$^{-1}$                  & 27 & 40 &  58 & $\pm 15$  & $\pm 38$\% \\
5HR     6 cm$^{-1}$                  & 20 & 27 &  37 & $\pm 8.7$ & $\pm 33$\% \\
5HR     9 cm$^{-1}$                  & 25 & 37 &  55 & $\pm 15$  & $\pm 41$\% \\
5PH   3.5 cm$^{-1}$                  & 32 & 50 &  77 & $\pm 23$  & $\pm 45$\% \\
5PH     6 cm$^{-1}$                  & 56 & 74 &  99 & $\pm 22$  & $\pm 30$\% \\
4     3.5 cm$^{-1}$                  & 54 & 71 &  93 & $\pm 19$  & $\pm 28$\% \\
4       6 cm$^{-1}$                  & 32 & 55 &  84 & $\pm 26$  & $\pm 47$\% \\
4       9 cm$^{-1}$                  & 35 & 54 &  80 & $\pm 22$  & $\pm 41$\% \\
5noMP\tablenotemark{e} { } 3.5 cm$^{-1}$& 33 & 44 &  58 & $\pm 12$  & $\pm 28$\% \\
5noMP\tablenotemark{e} { }   6 cm$^{-1}$& 41 & 51 &  63 & $\pm 11$  & $\pm 21$\% \\
\tableline
\end{tabular}
\end{center}
\end{table}

\begin{table}
\begin{center}
\caption{Numerical Values for Bandtemperature\tablenotemark{a}{ } 
from Likelihood Analyses Assuming a Flat Bandpower
Spectrum, for the Combined-Channel Data Sets}
\vspace{0.3truecm}
\label{tab:bandtemp2}
\tablenotetext{{\rm a}}{$\delta T_l$ in $\mu$K.}
\tablenotetext{{\rm b}}{Average absolute error in $\mu$K.} 
\tablenotetext{{\rm c}}{Average fractional error, 
 as a fraction of the central value.}
\tablenotetext{{\rm d}}{5MP does not have a $2\sigma$ HPD detection. The 
appropriate ET $2\sigma$ upper limit is 42 $\mu$K.}
\tablenotetext{{\rm e}}{Data from 5HR and 5PH.}
\tablenotetext{{\rm f}}{C(ombined) data excluding 5MP, i.e., data from
4ID, 4SH, 5HR, and 5PH.}
\tablenotetext{{\rm g}}{C(ombined) data including 5MP.}
\begin{tabular}{lccccc}
\tableline\tableline
Data Set & $-1\sigma$ & Peak & $+1\sigma$ &
 Ave. Abs. Err.\tablenotemark{b} & Ave. Fra. Err.\tablenotemark{c} \\
\tableline
4ID                    & 38 & 55  &  83 & $\pm23$  &  $\pm41$\% \\
4SH                    & 58 & 84  & 130 & $\pm34$  &  $\pm40$\% \\
5HR                    & 22 & 28  &  38 & $\pm8.4$ &  $\pm30$\% \\
5MP\tablenotemark{d}   &  0 & 2.1 &  17 & $\pm8.5$ & $\pm410$\% \\
5PH                    & 56 & 74  &  98 & $\pm21$  &  $\pm28$\% \\
4                      & 53 & 69  &  90 & $\pm19$  &  $\pm27$\% \\
5noMP\tablenotemark{e} & 42 & 51  &  63 & $\pm10$  &  $\pm20$\% \\
5                      & 37 & 43  &  52 & $\pm7.5$ &  $\pm17$\% \\
CnoMP\tablenotemark{f} & 49 & 58  &  67 & $\pm9.0$ &  $\pm16$\% \\
C\tablenotemark{g}     & 44 & 51  &  59 & $\pm7.2$ &  $\pm14$\% \\
\tableline
\end{tabular}
\end{center}
\end{table}

\begin{table}
\renewcommand{\arraystretch}{0.75}
\begin{center}
\caption{Numerical Values for $\qrmsps$ (in $\mu$K)\tablenotemark{a}}
\vspace{0.3truecm}
\label{tab:maqrmsps1}
\tablenotetext{{\rm a}}{For each model, the first of the three entries in
each of the last six columns is where the probability density
distribution function peaks. Ellipses as the lower entry in a vertical pair
denotes a non-detection; the upper entry then is the $2\sigma$
(97.7\% ET) upper limit.  For detections, the vertical pair of numbers are the
$\pm 1\sigma$ (68.3\% HPD) limits, except for DMR where
they are $\pm 2\sigma$ (95.5\% HPD).}
\tablenotetext{{\rm b}}{DMR values are from G\'orski et al. (1996,1998) and 
Stompor (1997).}
\begin{tabular}{cc|c|ccccc}
\tableline\tableline
Model & ($\Omega_0$, $h$, $\Omega_B h^2$) & DMR\tablenotemark{b} 
      & 4ID & 4SH & 5HR & 5MP & 5PH \\
\tableline
         O1 & (0.1, 0.75, 0.0125) & $21.0_{\,17.0}^{\,25.0}$ & $43_{\,29}^{\,64}$ & $61_{\,43}^{\,89}$ & $23_{\,18}^{\,31}$ & $ 0_{\,\cdots}^{\,32}$ & $52_{\,40}^{\,69}$\\
         O2 & (0.2, 0.65, 0.0175) & $24.1_{\,19.6}^{\,28.6}$ & $38_{\,26}^{\,56}$ & $54_{\,38}^{\,78}$ & $21_{\,16}^{\,28}$ & $ 0_{\,\cdots}^{\,28}$ & $45_{\,35}^{\,60}$\\
         O3 & (0.2, 0.70, 0.0125) & $24.1_{\,19.6}^{\,28.6}$ & $40_{\,27}^{\,59}$ & $57_{\,40}^{\,82}$ & $22_{\,17}^{\,29}$ & $ 0_{\,\cdots}^{\,30}$ & $48_{\,37}^{\,63}$\\
         O4 & (0.2, 0.75, 0.0075) & $24.1_{\,19.6}^{\,28.6}$ & $42_{\,29}^{\,63}$ & $60_{\,42}^{\,88}$ & $23_{\,18}^{\,31}$ & $ 0_{\,\cdots}^{\,31}$ & $51_{\,39}^{\,67}$\\
         O5 & (0.3, 0.60, 0.0175) & $23.4_{\,19.1}^{\,27.8}$ & $31_{\,21}^{\,46}$ & $44_{\,31}^{\,63}$ & $17_{\,13}^{\,23}$ & $ 0_{\,\cdots}^{\,23}$ & $37_{\,28}^{\,49}$\\
         O6 & (0.3, 0.65, 0.0125) & $23.4_{\,19.1}^{\,27.8}$ & $33_{\,23}^{\,49}$ & $47_{\,33}^{\,68}$ & $18_{\,14}^{\,24}$ & $ 0_{\,\cdots}^{\,24}$ & $39_{\,30}^{\,52}$\\
         O7 & (0.3, 0.70, 0.0075) & $23.4_{\,19.1}^{\,27.8}$ & $35_{\,24}^{\,52}$ & $50_{\,35}^{\,73}$ & $19_{\,15}^{\,26}$ & $ 0_{\,\cdots}^{\,26}$ & $42_{\,33}^{\,56}$\\
         O8 & (0.4, 0.60, 0.0175) & $21.2_{\,17.4}^{\,24.9}$ & $25_{\,17}^{\,37}$ & $36_{\,25}^{\,52}$ & $14_{\,11}^{\,19}$ & $ 0_{\,\cdots}^{\,19}$ & $30_{\,23}^{\,40}$\\
         O9 & (0.4, 0.65, 0.0125) & $21.2_{\,17.4}^{\,24.9}$ & $27_{\,18}^{\,40}$ & $38_{\,27}^{\,56}$ & $15_{\,11}^{\,20}$ & $ 0_{\,\cdots}^{\,20}$ & $32_{\,25}^{\,43}$\\
        O10 & (0.4, 0.70, 0.0075) & $21.2_{\,17.4}^{\,24.9}$ & $29_{\,20}^{\,43}$ & $41_{\,29}^{\,60}$ & $16_{\,12}^{\,22}$ & $ 0_{\,\cdots}^{\,21}$ & $35_{\,27}^{\,46}$\\
        O11 & (0.5, 0.55, 0.0175) & $18.5_{\,15.3}^{\,21.7}$ & $20_{\,13}^{\,29}$ & $28_{\,20}^{\,41}$ & $11_{\,8.5}^{\,15}$ & $ 0_{\,\cdots}^{\,14}$ & $24_{\,18}^{\,31}$\\
        O12 & (0.5, 0.60, 0.0125) & $18.5_{\,15.3}^{\,21.7}$ & $21_{\,15}^{\,32}$ & $30_{\,21}^{\,44}$ & $12_{\,9.1}^{\,16}$ & $ 0_{\,\cdots}^{\,16}$ & $26_{\,20}^{\,34}$\\
        O13 & (0.5, 0.65, 0.0075) & $18.5_{\,15.3}^{\,21.7}$ & $23_{\,16}^{\,34}$ & $33_{\,23}^{\,48}$ & $13_{\,9.8}^{\,17}$ & $ 0_{\,\cdots}^{\,17}$ & $28_{\,22}^{\,37}$\\
        O14 & (1.0, 0.50, 0.0125) & $17.8_{\,14.7}^{\,20.8}$ & $18_{\,12}^{\,27}$ & $27_{\,19}^{\,40}$ & $10_{\,7.9}^{\,14}$ & $ 0_{\,\cdots}^{\,14}$ & $23_{\,18}^{\,31}$\\
 $\Lambda$1 & (0.1, 0.90, 0.0125) & $25.5_{\,21.2}^{\,29.7}$ & $20_{\,14}^{\,30}$ & $30_{\,20}^{\,44}$ & $12_{\,8.8}^{\,16}$ & $ 0_{\,\cdots}^{\,15}$ & $26_{\,20}^{\,34}$\\
 $\Lambda$2 & (0.2, 0.80, 0.0075) & $23.3_{\,19.4}^{\,27.1}$ & $20_{\,14}^{\,30}$ & $30_{\,21}^{\,44}$ & $12_{\,8.8}^{\,16}$ & $ 0_{\,\cdots}^{\,15}$ & $26_{\,20}^{\,34}$\\
 $\Lambda$3 & (0.2, 0.75, 0.0125) & $23.1_{\,19.2}^{\,27.0}$ & $19_{\,13}^{\,28}$ & $27_{\,19}^{\,41}$ & $11_{\,8.1}^{\,14}$ & $ 0_{\,\cdots}^{\,14}$ & $24_{\,18}^{\,31}$\\
 $\Lambda$4 & (0.2, 0.70, 0.0175) & $23.0_{\,19.1}^{\,26.8}$ & $17_{\,12}^{\,26}$ & $25_{\,17}^{\,38}$ & $10_{\,7.6}^{\,13}$ & $ 0_{\,\cdots}^{\,13}$ & $22_{\,17}^{\,29}$\\
 $\Lambda$5 & (0.3, 0.70, 0.0075) & $21.0_{\,17.5}^{\,24.5}$ & $19_{\,13}^{\,28}$ & $28_{\,19}^{\,42}$ & $11_{\,8.2}^{\,15}$ & $ 0_{\,\cdots}^{\,14}$ & $24_{\,18}^{\,32}$\\
 $\Lambda$6 & (0.3, 0.65, 0.0125) & $21.0_{\,17.4}^{\,24.5}$ & $17_{\,12}^{\,26}$ & $26_{\,18}^{\,38}$ & $10_{\,7.6}^{\,13}$ & $ 0_{\,\cdots}^{\,13}$ & $22_{\,17}^{\,29}$\\
 $\Lambda$7 & (0.3, 0.60, 0.0175) & $20.9_{\,17.4}^{\,24.4}$ & $16_{\,11}^{\,24}$ & $24_{\,16}^{\,35}$ & $9.3_{\,7.1}^{\,12}$ & $ 0_{\,\cdots}^{\,12}$ & $20_{\,16}^{\,27}$\\
 $\Lambda$8 & (0.4, 0.65, 0.0075) & $19.4_{\,16.1}^{\,22.7}$ & $18_{\,13}^{\,28}$ & $27_{\,19}^{\,41}$ & $11_{\,8.0}^{\,14}$ & $ 0_{\,\cdots}^{\,14}$ & $23_{\,18}^{\,31}$\\
 $\Lambda$9 & (0.4, 0.60, 0.0125) & $19.4_{\,16.1}^{\,22.7}$ & $17_{\,12}^{\,25}$ & $25_{\,17}^{\,37}$ & $9.7_{\,7.4}^{\,13}$ & $ 0_{\,\cdots}^{\,13}$ & $21_{\,16}^{\,28}$\\
$\Lambda$10 & (0.4, 0.55, 0.0175) & $19.4_{\,16.1}^{\,22.7}$ & $15_{\,11}^{\,23}$ & $23_{\,16}^{\,34}$ & $9.0_{\,6.8}^{\,12}$ & $ 0_{\,\cdots}^{\,12}$ & $19_{\,15}^{\,26}$\\
$\Lambda$11 & (0.5, 0.60, 0.0125) & $18.5_{\,15.3}^{\,21.6}$ & $17_{\,12}^{\,26}$ & $25_{\,17}^{\,38}$ & $9.8_{\,7.5}^{\,13}$ & $ 0_{\,\cdots}^{\,13}$ & $22_{\,17}^{\,29}$\\
Flat        & $\cdots$            & $18.6_{\,15.4}^{\,21.8}$ & $36_{\,24}^{\,53}$ & $54_{\,37}^{\,81}$ & $18_{\,14}^{\,25}$ & $1.3_{\,\cdots}^{\,27}$ & $47_{\,36}^{\,63}$\\

\tableline
\end{tabular}
\end{center}
\end{table}

\begin{table}
\renewcommand{\arraystretch}{0.8}
\begin{center}
\caption{Numerical Values for $\qrmsps$ (in $\mu$K)\tablenotemark{a}}
\vspace{0.3truecm}
\label{tab:maqrmsps2}
\tablenotetext{{\rm a}}{Conventions are the same as for Table 6.}
\tablenotetext{{\rm b}}{Data from 5HR and 5PH.}
\tablenotetext{{\rm c}}{C(ombined) data excluding 5MP.}
\tablenotetext{{\rm d}}{C(ombined) data including 5MP.}
\begin{tabular}{cc|c|ccccc}
\tableline\tableline
Model & ($\Omega_0$, $h$, $\Omega_B h^2$) & DMR & 4 & 5noMP\tablenotemark{b} 
& 5 & CnoMP\tablenotemark{c} & C\tablenotemark{d} \\
\tableline
         O1 & (0.1, 0.75, 0.0125) & $21.0_{\,17.0}^{\,25.0}$ & $52_{\,40}^{\,67}$ & $38_{\,31}^{\,46}$ & $32_{\,27}^{\,38}$ & $43_{\,37}^{\,50}$ & $38_{\,33}^{\,44}$\\
         O2 & (0.2, 0.65, 0.0175) & $24.1_{\,19.6}^{\,28.6}$ & $46_{\,36}^{\,59}$ & $33_{\,28}^{\,40}$ & $28_{\,24}^{\,34}$ & $38_{\,32}^{\,44}$ & $33_{\,29}^{\,38}$\\
         O3 & (0.2, 0.70, 0.0125) & $24.1_{\,19.6}^{\,28.6}$ & $48_{\,37}^{\,62}$ & $35_{\,29}^{\,42}$ & $30_{\,25}^{\,35}$ & $40_{\,34}^{\,46}$ & $35_{\,31}^{\,40}$\\
         O4 & (0.2, 0.75, 0.0075) & $24.1_{\,19.6}^{\,28.6}$ & $51_{\,40}^{\,66}$ & $37_{\,31}^{\,45}$ & $32_{\,27}^{\,38}$ & $42_{\,36}^{\,49}$ & $37_{\,32}^{\,43}$\\
         O5 & (0.3, 0.60, 0.0175) & $23.4_{\,19.1}^{\,27.8}$ & $37_{\,29}^{\,48}$ & $27_{\,23}^{\,33}$ & $23_{\,20}^{\,28}$ & $31_{\,26}^{\,36}$ & $27_{\,24}^{\,31}$\\
         O6 & (0.3, 0.65, 0.0125) & $23.4_{\,19.1}^{\,27.8}$ & $40_{\,31}^{\,51}$ & $29_{\,24}^{\,35}$ & $25_{\,21}^{\,29}$ & $33_{\,28}^{\,38}$ & $29_{\,25}^{\,33}$\\
         O7 & (0.3, 0.70, 0.0075) & $23.4_{\,19.1}^{\,27.8}$ & $42_{\,33}^{\,55}$ & $31_{\,26}^{\,38}$ & $27_{\,23}^{\,31}$ & $35_{\,30}^{\,41}$ & $31_{\,27}^{\,36}$\\
         O8 & (0.4, 0.60, 0.0175) & $21.2_{\,17.4}^{\,24.9}$ & $30_{\,24}^{\,39}$ & $22_{\,18}^{\,27}$ & $19_{\,16}^{\,22}$ & $25_{\,22}^{\,29}$ & $22_{\,19}^{\,26}$\\
         O9 & (0.4, 0.65, 0.0125) & $21.2_{\,17.4}^{\,24.9}$ & $32_{\,25}^{\,42}$ & $24_{\,20}^{\,29}$ & $20_{\,17}^{\,24}$ & $27_{\,23}^{\,31}$ & $24_{\,21}^{\,27}$\\
        O10 & (0.4, 0.70, 0.0075) & $21.2_{\,17.4}^{\,24.9}$ & $35_{\,27}^{\,46}$ & $26_{\,21}^{\,31}$ & $22_{\,19}^{\,26}$ & $29_{\,25}^{\,34}$ & $26_{\,22}^{\,30}$\\
        O11 & (0.5, 0.55, 0.0175) & $18.5_{\,15.3}^{\,21.7}$ & $24_{\,18}^{\,31}$ & $17_{\,15}^{\,21}$ & $15_{\,13}^{\,18}$ & $20_{\,17}^{\,23}$ & $17_{\,15}^{\,20}$\\
        O12 & (0.5, 0.60, 0.0125) & $18.5_{\,15.3}^{\,21.7}$ & $26_{\,20}^{\,33}$ & $19_{\,16}^{\,23}$ & $16_{\,14}^{\,19}$ & $21_{\,18}^{\,25}$ & $19_{\,16}^{\,22}$\\
        O13 & (0.5, 0.65, 0.0075) & $18.5_{\,15.3}^{\,21.7}$ & $28_{\,22}^{\,36}$ & $20_{\,17}^{\,25}$ & $18_{\,15}^{\,21}$ & $23_{\,20}^{\,27}$ & $20_{\,18}^{\,24}$\\
        O14 & (1.0, 0.50, 0.0125) & $17.8_{\,14.7}^{\,20.8}$ & $22_{\,17}^{\,29}$ & $17_{\,14}^{\,21}$ & $14_{\,12}^{\,17}$ & $19_{\,16}^{\,22}$ & $17_{\,14}^{\,19}$\\
 $\Lambda$1 & (0.1, 0.90, 0.0125) & $25.5_{\,21.2}^{\,29.7}$ & $25_{\,19}^{\,32}$ & $19_{\,16}^{\,23}$ & $16_{\,14}^{\,19}$ & $21_{\,18}^{\,24}$ & $18_{\,16}^{\,21}$\\
 $\Lambda$2 & (0.2, 0.80, 0.0075) & $23.3_{\,19.4}^{\,27.1}$ & $25_{\,19}^{\,32}$ & $19_{\,16}^{\,23}$ & $16_{\,14}^{\,19}$ & $21_{\,18}^{\,24}$ & $18_{\,16}^{\,21}$\\
 $\Lambda$3 & (0.2, 0.75, 0.0125) & $23.1_{\,19.2}^{\,27.0}$ & $23_{\,18}^{\,30}$ & $17_{\,14}^{\,21}$ & $15_{\,12}^{\,17}$ & $19_{\,17}^{\,22}$ & $17_{\,15}^{\,20}$\\
 $\Lambda$4 & (0.2, 0.70, 0.0175) & $23.0_{\,19.1}^{\,26.8}$ & $21_{\,16}^{\,28}$ & $16_{\,13}^{\,19}$ & $14_{\,12}^{\,16}$ & $18_{\,15}^{\,21}$ & $16_{\,14}^{\,18}$\\
 $\Lambda$5 & (0.3, 0.70, 0.0075) & $21.0_{\,17.5}^{\,24.5}$ & $23_{\,18}^{\,30}$ & $18_{\,15}^{\,21}$ & $15_{\,13}^{\,18}$ & $20_{\,17}^{\,23}$ & $17_{\,15}^{\,20}$\\
 $\Lambda$6 & (0.3, 0.65, 0.0125) & $21.0_{\,17.4}^{\,24.5}$ & $21_{\,16}^{\,28}$ & $16_{\,13}^{\,20}$ & $14_{\,12}^{\,16}$ & $18_{\,15}^{\,21}$ & $16_{\,14}^{\,18}$\\
 $\Lambda$7 & (0.3, 0.60, 0.0175) & $20.9_{\,17.4}^{\,24.4}$ & $20_{\,15}^{\,26}$ & $15_{\,12}^{\,18}$ & $13_{\,11}^{\,15}$ & $16_{\,14}^{\,19}$ & $15_{\,13}^{\,17}$\\
 $\Lambda$8 & (0.4, 0.65, 0.0075) & $19.4_{\,16.1}^{\,22.7}$ & $23_{\,17}^{\,30}$ & $17_{\,14}^{\,21}$ & $15_{\,12}^{\,17}$ & $19_{\,16}^{\,22}$ & $17_{\,15}^{\,19}$\\
 $\Lambda$9 & (0.4, 0.60, 0.0125) & $19.4_{\,16.1}^{\,22.7}$ & $21_{\,16}^{\,27}$ & $16_{\,13}^{\,19}$ & $13_{\,11}^{\,16}$ & $17_{\,15}^{\,20}$ & $15_{\,13}^{\,18}$\\
$\Lambda$10 & (0.4, 0.55, 0.0175) & $19.4_{\,16.1}^{\,22.7}$ & $19_{\,15}^{\,25}$ & $14_{\,12}^{\,17}$ & $12_{\,10}^{\,14}$ & $16_{\,14}^{\,19}$ & $14_{\,12}^{\,16}$\\
$\Lambda$11 & (0.5, 0.60, 0.0125) & $18.5_{\,15.3}^{\,21.6}$ & $21_{\,16}^{\,28}$ & $16_{\,13}^{\,19}$ & $14_{\,12}^{\,16}$ & $18_{\,15}^{\,21}$ & $16_{\,14}^{\,18}$\\
Flat        & $\cdots$            & $18.6_{\,15.4}^{\,21.8}$ & $44_{\,34}^{\,58}$ & $33_{\,27}^{\,41}$ & $28_{\,24}^{\,33}$ & $37_{\,32}^{\,43}$ & $33_{\,28}^{\,38}$\\

\tableline
\end{tabular}
\end{center}
\end{table}

\begin{table}
\renewcommand{\arraystretch}{0.8}
\begin{center}
\caption{Renormalized Maximum Values of the Probability Density Distribution
Functions\tablenotemark{a}}
\vspace{0.3truecm}
\label{tab:mamaxlh}
\tablenotetext{{\rm a}}{Renormalized such that it is unity for the 
  ``realistic'' model with the highest maximum value of the probability density
  distribution function for the data set. The last line of the table
  gives this highest maximum likelihood value when the normalization is set
  such that $L(Q_{\rm rms-PS} = 0~\mu{\rm K}) = 1$.}
\begin{tabular}{c|cccccccccc}
\tableline\tableline
Model & 4ID & 4SH & 5HR & 5MP & 5PH & 4 & 5noMP & 5 & CnoMP & C \\
\tableline
         O1 & 0.92 & 0.90 & 1.0  & 1.0  & 0.66 & 0.95 & 0.72 & 0.89 & 0.75 & 0.88 \\
         O2 & 0.87 & 0.98 & 0.70 & 1.0  & 0.66 & 0.99 & 0.61 & 0.76 & 0.67 & 0.78 \\
         O3 & 0.88 & 0.96 & 0.79 & 1.0  & 0.66 & 0.98 & 0.65 & 0.81 & 0.70 & 0.82 \\
         O4 & 0.90 & 0.93 & 0.90 & 1.0  & 0.66 & 0.97 & 0.69 & 0.85 & 0.73 & 0.85 \\
         O5 & 0.85 & 1.0  & 0.61 & 1.0  & 0.69 & 1.0  & 0.61 & 0.73 & 0.67 & 0.76 \\
         O6 & 0.87 & 0.98 & 0.70 & 1.0  & 0.70 & 0.99 & 0.66 & 0.79 & 0.72 & 0.82 \\
         O7 & 0.88 & 0.96 & 0.80 & 1.0  & 0.70 & 0.98 & 0.70 & 0.85 & 0.77 & 0.87 \\
         O8 & 0.85 & 0.98 & 0.60 & 1.0  & 0.76 & 0.99 & 0.67 & 0.77 & 0.75 & 0.81 \\
         O9 & 0.87 & 0.96 & 0.69 & 1.0  & 0.78 & 0.98 & 0.74 & 0.85 & 0.82 & 0.88 \\
        O10 & 0.88 & 0.95 & 0.79 & 1.0  & 0.79 & 0.97 & 0.79 & 0.92 & 0.87 & 0.95 \\
        O11 & 0.86 & 0.96 & 0.58 & 1.0  & 0.82 & 0.97 & 0.72 & 0.78 & 0.80 & 0.83 \\
        O12 & 0.88 & 0.94 & 0.66 & 1.0  & 0.85 & 0.96 & 0.79 & 0.87 & 0.88 & 0.92 \\
        O13 & 0.89 & 0.93 & 0.75 & 1.0  & 0.86 & 0.96 & 0.86 & 0.96 & 0.95 & 1.0  \\
        O14 & 0.99 & 0.71 & 0.83 & 1.0  & 1.0  & 0.79 & 1.0  & 1.0  & 1.0  & 1.0  \\
 $\Lambda$1 & 0.99 & 0.70 & 0.74 & 1.0  & 0.97 & 0.77 & 0.91 & 0.86 & 0.90 & 0.86 \\
 $\Lambda$2 & 1.0  & 0.69 & 0.78 & 1.0  & 0.96 & 0.77 & 0.92 & 0.90 & 0.91 & 0.89 \\
 $\Lambda$3 & 0.99 & 0.71 & 0.74 & 1.0  & 0.98 & 0.78 & 0.92 & 0.88 & 0.92 & 0.89 \\
 $\Lambda$4 & 0.96 & 0.73 & 0.67 & 1.0  & 0.99 & 0.79 & 0.90 & 0.86 & 0.92 & 0.87 \\
 $\Lambda$5 & 0.99 & 0.71 & 0.77 & 1.0  & 0.97 & 0.79 & 0.93 & 0.91 & 0.92 & 0.91 \\
 $\Lambda$6 & 0.98 & 0.72 & 0.72 & 1.0  & 0.99 & 0.79 & 0.92 & 0.90 & 0.93 & 0.90 \\
 $\Lambda$7 & 0.96 & 0.75 & 0.65 & 1.0  & 1.00 & 0.81 & 0.90 & 0.86 & 0.92 & 0.88 \\
 $\Lambda$8 & 0.99 & 0.71 & 0.78 & 1.0  & 0.97 & 0.79 & 0.94 & 0.93 & 0.94 & 0.93 \\
 $\Lambda$9 & 0.98 & 0.73 & 0.73 & 1.0  & 0.99 & 0.79 & 0.94 & 0.91 & 0.94 & 0.92 \\
$\Lambda$10 & 0.96 & 0.76 & 0.65 & 1.0  & 1.0  & 0.81 & 0.91 & 0.87 & 0.93 & 0.89 \\
$\Lambda$11 & 0.99 & 0.71 & 0.78 & 1.0  & 0.99 & 0.78 & 0.96 & 0.94 & 0.96 & 0.94 \\
       Flat & 1.1 & 0.61 & 2.6 & 1.0 & 0.58 & 0.71 & 0.78 & 0.78 & 0.68 & 0.66 \\
\tableline
\dots & $ 1 \cdot 10^{ 6} $  & $ 5 \cdot 10^{12} $  & $ 3 \cdot 10^{19} $  & $ 1 \cdot 10^{ 0} $  & $ 7 \cdot 10^{34} $  & $ 4 \cdot 10^{18} $  & $ 2 \cdot 10^{53} $  & $ 6 \cdot 10^{51} $  & $ 4 \cdot 10^{71} $  & $ 6 \cdot 10^{69} $  \\

\tableline
\end{tabular}
\end{center}
\end{table}

\begin{table}
\renewcommand{\arraystretch}{0.8}
\begin{center}
\caption{Renormalized Marginal Values of the Probability Density Distribution
Functions\tablenotemark{a}}
\vspace{0.3truecm}
\label{tab:mamarglh}
\tablenotetext{{\rm a}}{Renormalized such that it is unity for the 
  ``realistic'' model
  with the highest marginal probability density distribution function
  value for the data set. The last line of the table gives the
  marginal value for the model with highest marginal probability
  density distribution function value when the likelihoods are
  normalized such that $L(Q_{\rm rms-PS} = 0\ \mu{\rm K}) = 1$.}
\begin{tabular}{c|cccccccccc}
\tableline\tableline
Model & 4ID & 4SH & 5HR & 5MP & 5PH & 4 & 5noMP & 5 & CnoMP & C \\
\tableline
         O1 & 1.0  & 0.99 & 1.0  & 1.0  & 1.0  & 1.0  & 1.0  & 1.0  & 1.0  & 1.0  \\
         O2 & 0.83 & 0.93 & 0.63 & 0.88 & 0.86 & 0.91 & 0.74 & 0.75 & 0.77 & 0.78 \\
         O3 & 0.89 & 0.96 & 0.74 & 0.93 & 0.92 & 0.95 & 0.84 & 0.84 & 0.86 & 0.86 \\
         O4 & 0.96 & 1.0  & 0.89 & 0.98 & 0.97 & 0.99 & 0.94 & 0.94 & 0.95 & 0.95 \\
         O5 & 0.67 & 0.77 & 0.45 & 0.71 & 0.73 & 0.75 & 0.60 & 0.58 & 0.64 & 0.62 \\
         O6 & 0.72 & 0.81 & 0.55 & 0.76 & 0.80 & 0.79 & 0.70 & 0.67 & 0.73 & 0.71 \\
         O7 & 0.79 & 0.86 & 0.67 & 0.82 & 0.86 & 0.84 & 0.80 & 0.78 & 0.83 & 0.81 \\
         O8 & 0.55 & 0.62 & 0.37 & 0.58 & 0.65 & 0.60 & 0.55 & 0.50 & 0.58 & 0.53 \\
         O9 & 0.60 & 0.66 & 0.45 & 0.62 & 0.73 & 0.64 & 0.64 & 0.59 & 0.68 & 0.63 \\
        O10 & 0.65 & 0.71 & 0.54 & 0.67 & 0.80 & 0.69 & 0.75 & 0.70 & 0.79 & 0.73 \\
        O11 & 0.43 & 0.49 & 0.28 & 0.45 & 0.56 & 0.47 & 0.46 & 0.40 & 0.49 & 0.43 \\
        O12 & 0.48 & 0.52 & 0.34 & 0.49 & 0.63 & 0.50 & 0.56 & 0.49 & 0.58 & 0.52 \\
        O13 & 0.52 & 0.56 & 0.42 & 0.53 & 0.70 & 0.54 & 0.65 & 0.58 & 0.68 & 0.61 \\
        O14 & 0.47 & 0.36 & 0.38 & 0.42 & 0.67 & 0.37 & 0.63 & 0.50 & 0.59 & 0.50 \\
 $\Lambda$1 & 0.51 & 0.39 & 0.38 & 0.46 & 0.73 & 0.39 & 0.64 & 0.48 & 0.59 & 0.48 \\
 $\Lambda$2 & 0.52 & 0.39 & 0.40 & 0.47 & 0.72 & 0.40 & 0.64 & 0.49 & 0.59 & 0.49 \\
 $\Lambda$3 & 0.47 & 0.37 & 0.35 & 0.43 & 0.68 & 0.37 & 0.59 & 0.45 & 0.55 & 0.45 \\
 $\Lambda$4 & 0.43 & 0.35 & 0.29 & 0.40 & 0.63 & 0.35 & 0.54 & 0.40 & 0.51 & 0.41 \\
 $\Lambda$5 & 0.48 & 0.38 & 0.36 & 0.44 & 0.68 & 0.38 & 0.61 & 0.47 & 0.57 & 0.47 \\
 $\Lambda$6 & 0.44 & 0.35 & 0.31 & 0.40 & 0.64 & 0.35 & 0.56 & 0.43 & 0.53 & 0.43 \\
 $\Lambda$7 & 0.40 & 0.33 & 0.26 & 0.37 & 0.59 & 0.33 & 0.50 & 0.37 & 0.48 & 0.38 \\
 $\Lambda$8 & 0.47 & 0.37 & 0.36 & 0.43 & 0.66 & 0.37 & 0.60 & 0.47 & 0.56 & 0.47 \\
 $\Lambda$9 & 0.43 & 0.34 & 0.31 & 0.39 & 0.62 & 0.34 & 0.54 & 0.42 & 0.52 & 0.42 \\
$\Lambda$10 & 0.38 & 0.32 & 0.26 & 0.36 & 0.57 & 0.32 & 0.49 & 0.37 & 0.47 & 0.38 \\
$\Lambda$11 & 0.44 & 0.34 & 0.33 & 0.40 & 0.63 & 0.34 & 0.57 & 0.44 & 0.53 & 0.44 \\
       Flat & 1.0 & 0.62 & 2.1 & 0.84 & 0.82 & 0.66 & 0.98 & 0.78 & 0.81 & 0.66 \\
\tableline
\dots & $ 5 \cdot 10^{ 7} $  & $ 3 \cdot 10^{14} $  & $ 5 \cdot 10^{20} $  & $ 2 \cdot 10^{ 1} $  & $ 2 \cdot 10^{36} $  & $ 1 \cdot 10^{20} $  & $ 2 \cdot 10^{54} $  & $ 7 \cdot 10^{52} $  & $ 5 \cdot 10^{72} $  & $ 7 \cdot 10^{70} $  \\

\tableline
\end{tabular}
\end{center}
\end{table}

\clearpage
\centerline{FIGURE CAPTIONS}

\begin{figure}[htbp]
  \caption{CMB anisotropy multipole moments $l(l+1)C_l/(2\pi )\times
    10^{10}$ (broken lines, scale on left axis) as a function of
    multipole $l$, to $l = 600$, for selected models O1, O11, O14,
    $\Lambda$2, $\Lambda$10, and Flat, normalized to the DMR maps
    (G\'orski et al. 1996,1998; Stompor 1997). See Table 6 for model-parameter 
    values. Also shown are the four MAX individual-channel, zero-lag, nominal 
    beamwidth window functions $W_l$ (solid lines, scale on right axis):
    MAX 4 6/9 cm$^{-1}$; MAX 4 3.5 cm$^{-1}$; MAX 5 6/9 cm$^{-1}$;
    and MAX 5 3.5 cm$^{-1}$ with the $W_l$ peak moving from left to 
    right (the MAX 4 3.5 cm$^{-1}$ and MAX 5 6/9 cm$^{-1}$ window 
    functions overlap).}
\label{fig:spectra}\end{figure}

\begin{figure}[htbp]
  \caption{$(\delta T_{\rm rms}{}^2)_l$ [$= T_0{}^2 (2l+1) C_l W_l/(4\pi)$,
    where $T_0$ is the CMB temperature now] as a function of $l$, to $l = 400$, 
    for (upper panel) the MAX 4 6/9 cm$^{-1}$ $W_l$, and for (lower panel) the
    MAX 5 3.5 cm$^{-1}$ $W_l$, for the selected models shown in Fig. 1, 
    normalized to the DMR data.  See Tables 2, 6, and 7 for numerical values. 
    Note that the peak sensitivity of an individual-channel window function
    corresponds to a different angular scale in each of the models.}
\label{fig:dtrms}\end{figure}
 
\begin{figure}[htbp]
  \caption{Individual-channel MAX data (thermodynamic temperature).}
\label{fig:data}\end{figure}

\begin{figure}[htbp]
  \caption{Likelihood functions, as a function of $Q_{\rm rms-PS}$
    for the 6 selected models (O1, O11, O14, $\Lambda$2, $\Lambda$10, and 
    Flat) of Fig.  1 (line styles are identical to those used in Figs. 1 
    and 2), for the combined-channel MAX data sets. See Table 2 
    for model-parameter values, and 
    Tables 6 -- 9 for numerical values derived from the corresponding   
    probability density distribution functions.}
\label{fig:likenoms}\end{figure}

\clearpage

\begin{center}
  \leavevmode
  \epsfxsize=6.5truein
  \epsfbox{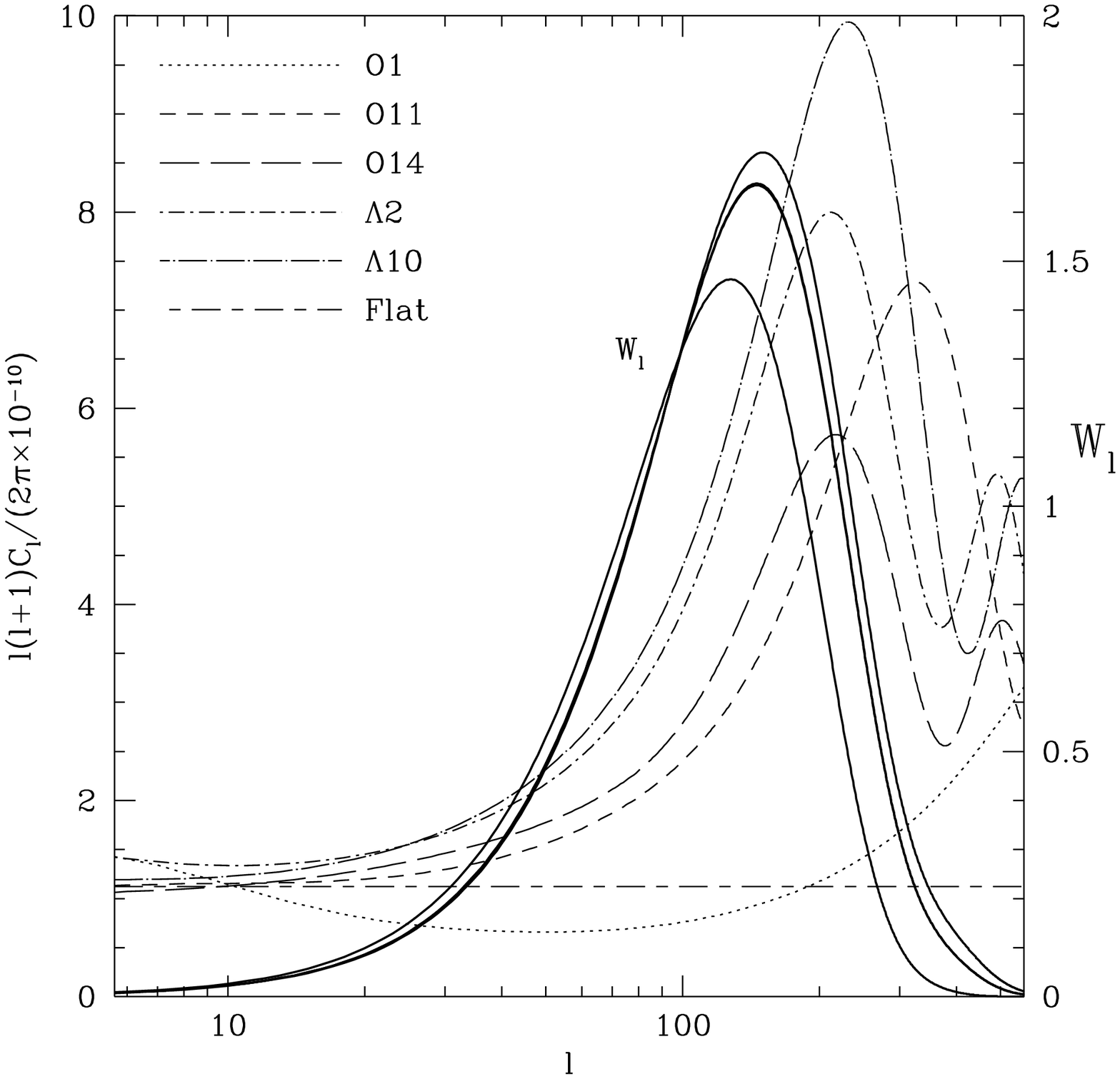}
\end{center}
\vfill
Figure~\ref{fig:spectra}

\clearpage

\begin{center}
  \leavevmode
  \epsfxsize=6.5truein
  \epsfbox{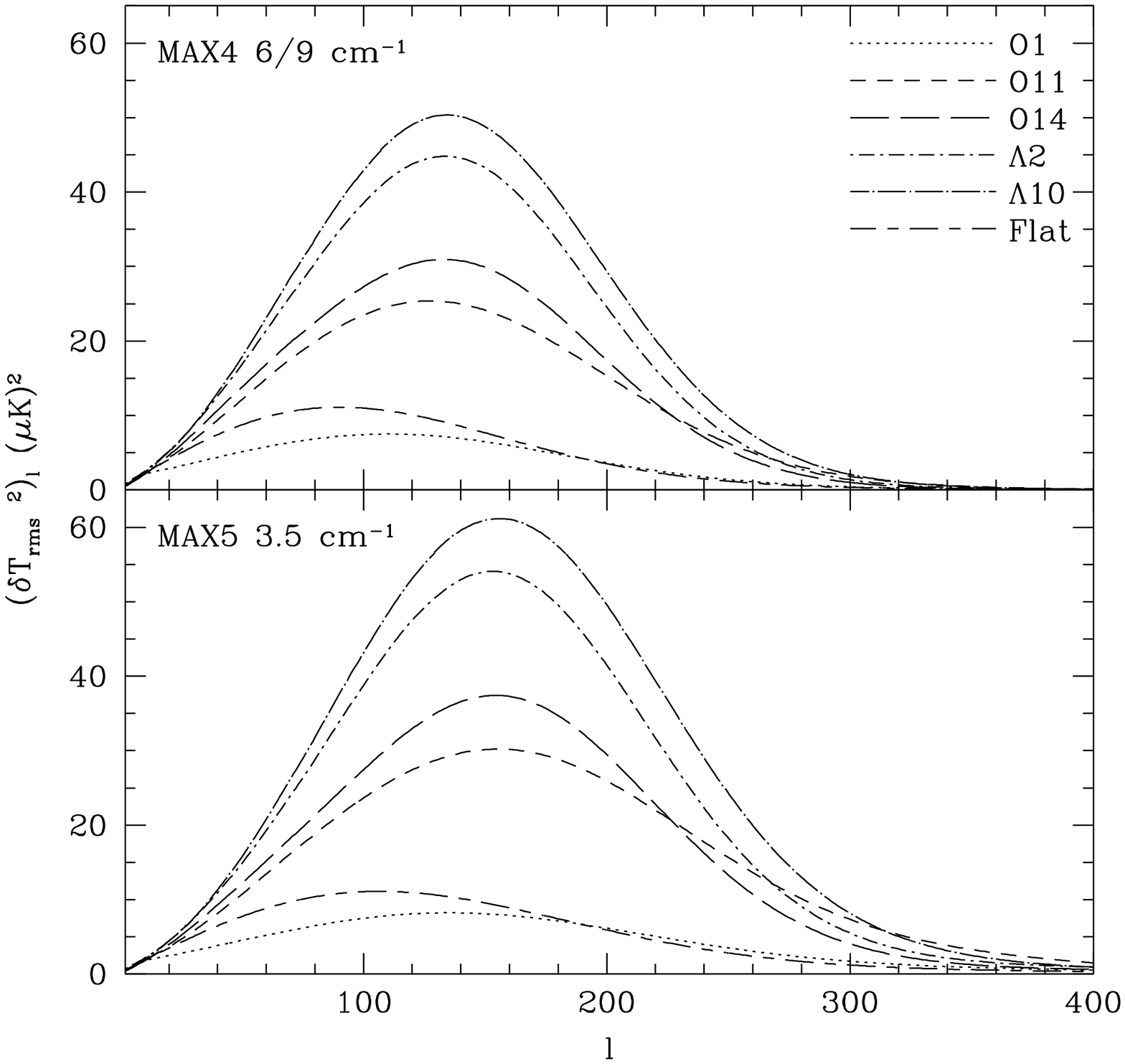}
\end{center}
\vfill
Figure~\ref{fig:dtrms}

\clearpage

\begin{center}
  \leavevmode
  \epsfxsize=6.5truein
  \epsfbox{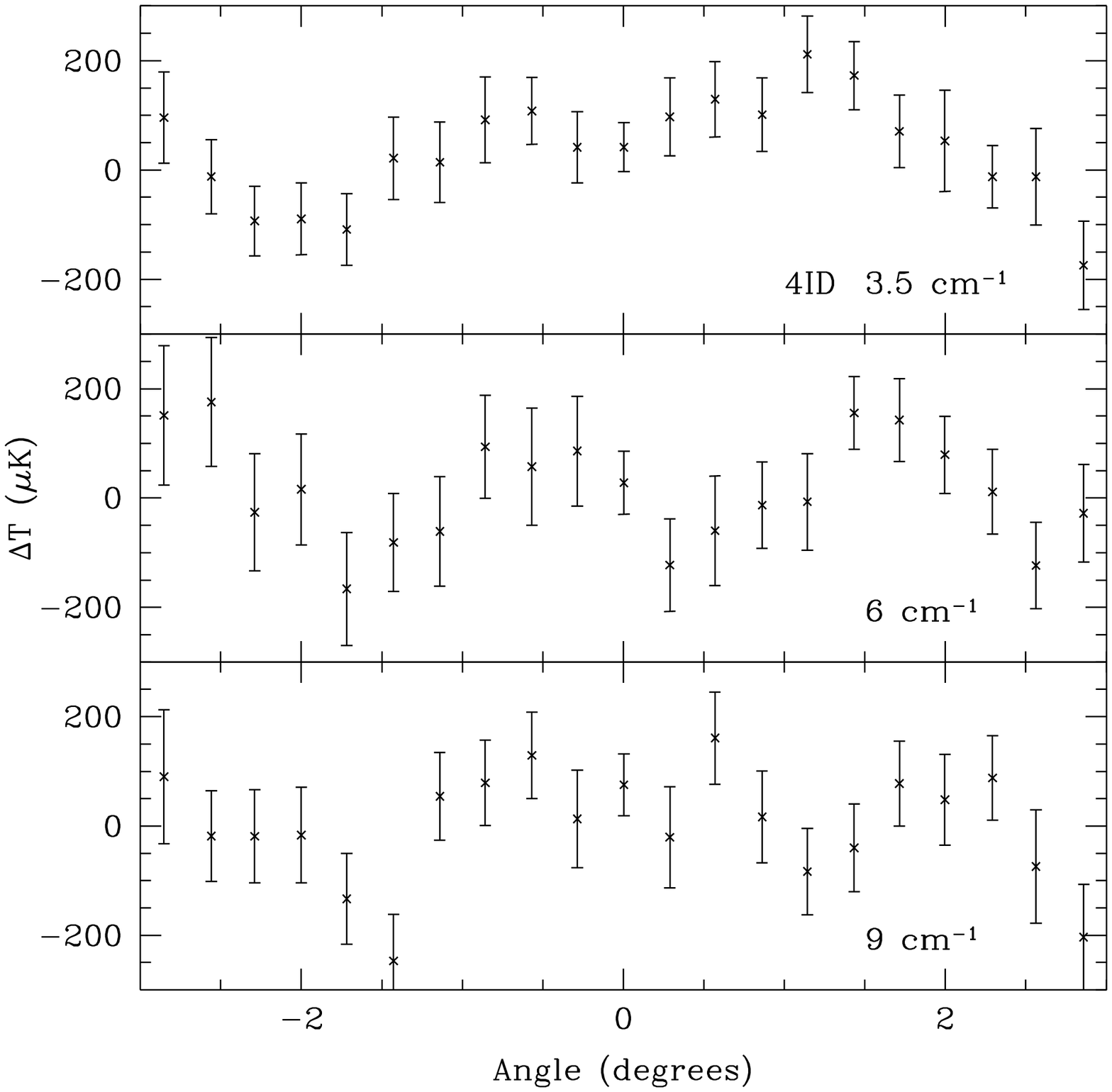}
\end{center}
\vfill
Figure~\ref{fig:data}a

\clearpage

\begin{center}
  \leavevmode
  \epsfxsize=6.5truein
  \epsfbox{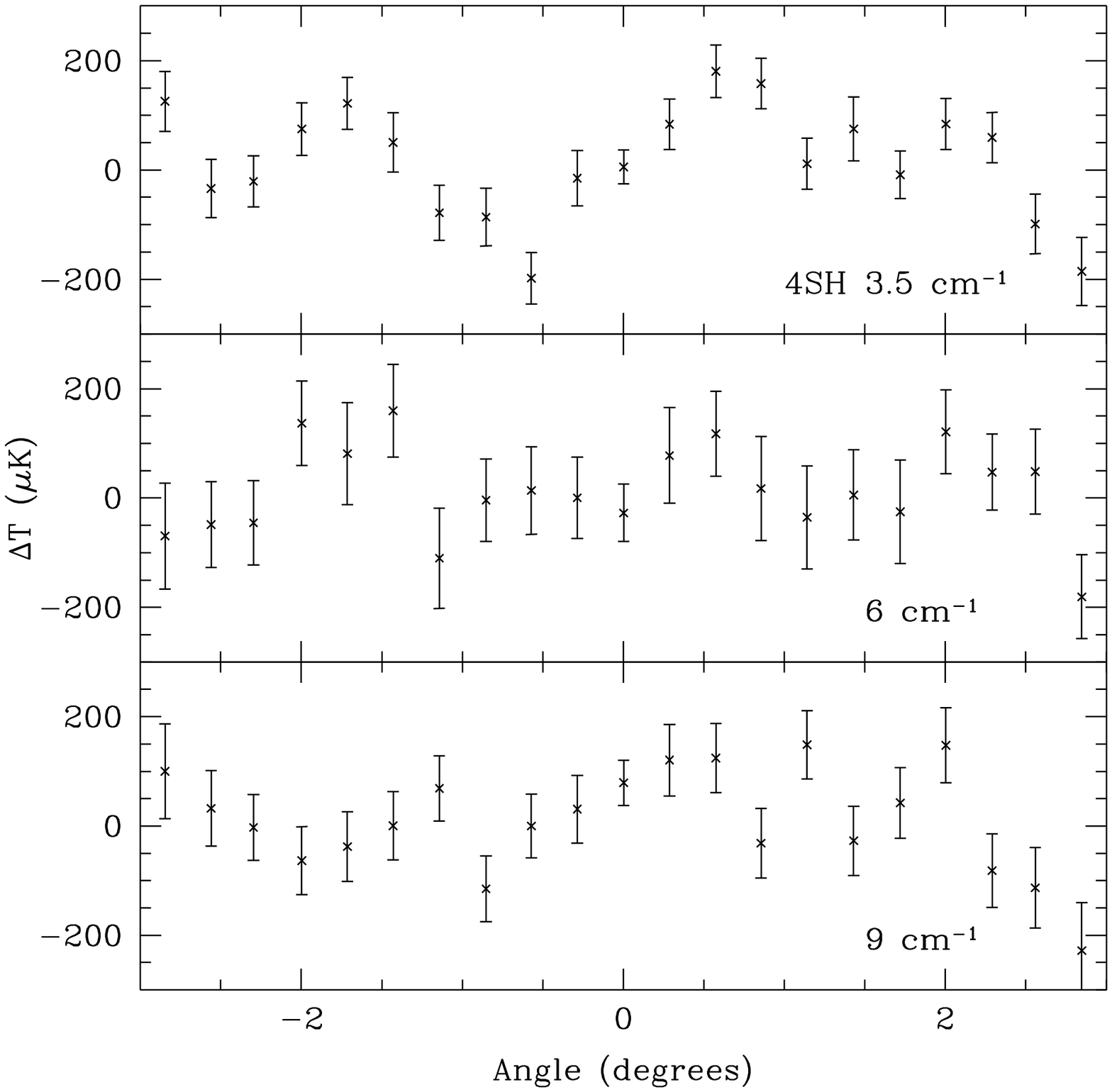}
\end{center}
\vfill
Figure~\ref{fig:data}b

\clearpage

\begin{center}
  \leavevmode
  \epsfxsize=6.5truein
  \epsfbox{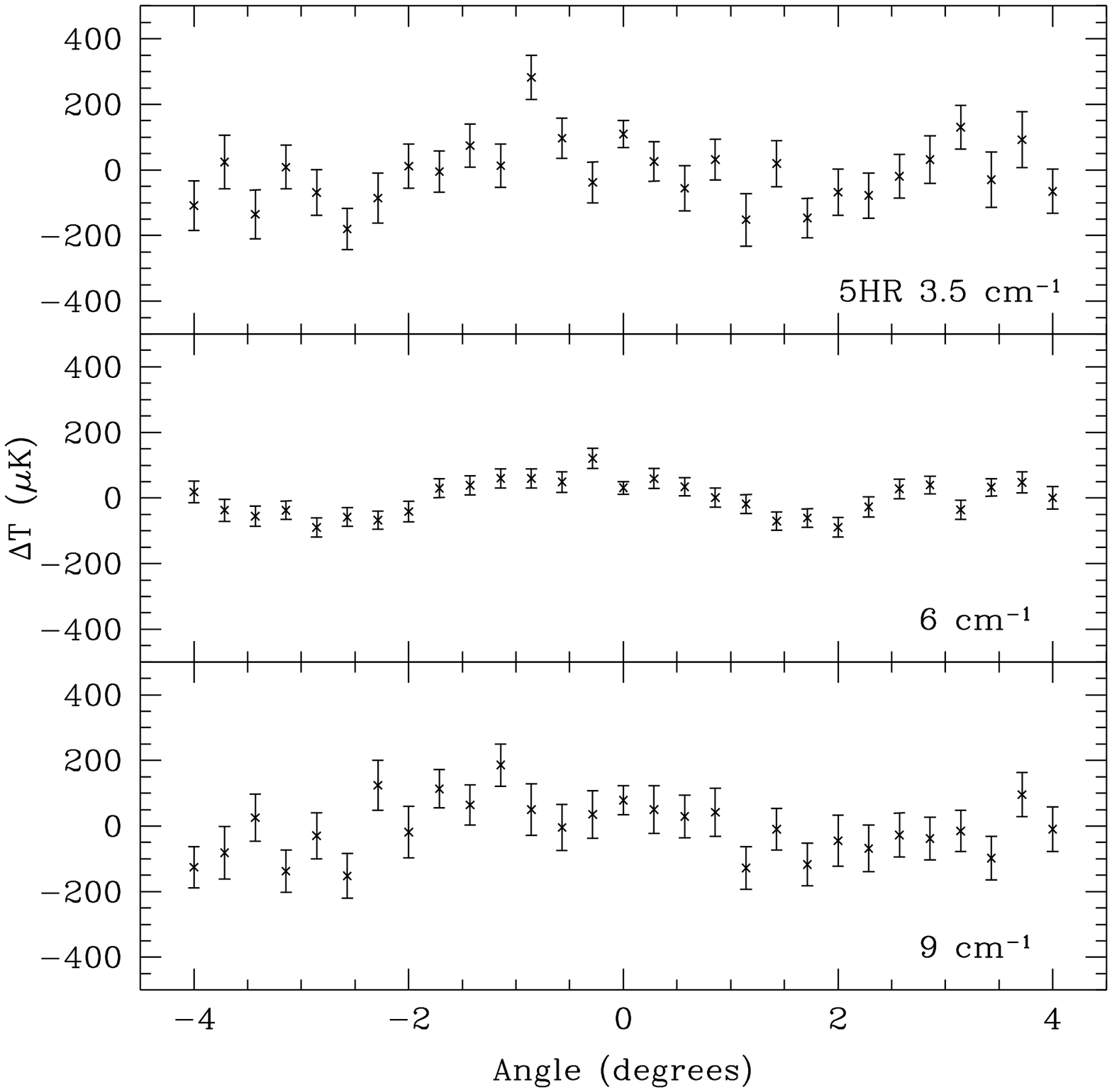}
\end{center}
\vfill
Figure~\ref{fig:data}c

\clearpage

\begin{center}
  \leavevmode
  \epsfxsize=6.5truein
  \epsfbox{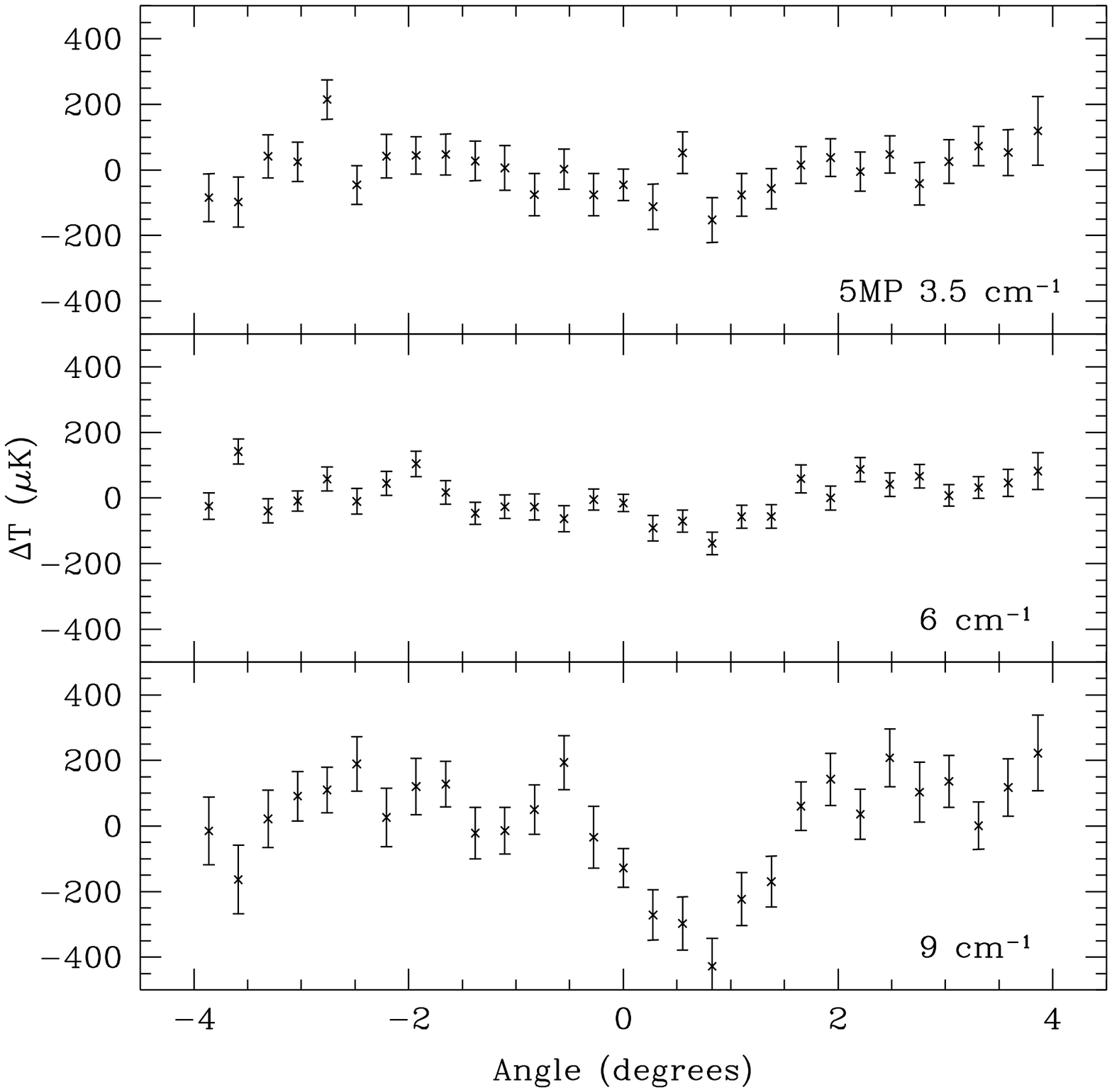}
\end{center}
\vfill
Figure~\ref{fig:data}d

\clearpage

\begin{center}
  \leavevmode
  \epsfxsize=6.5truein
  \epsfbox{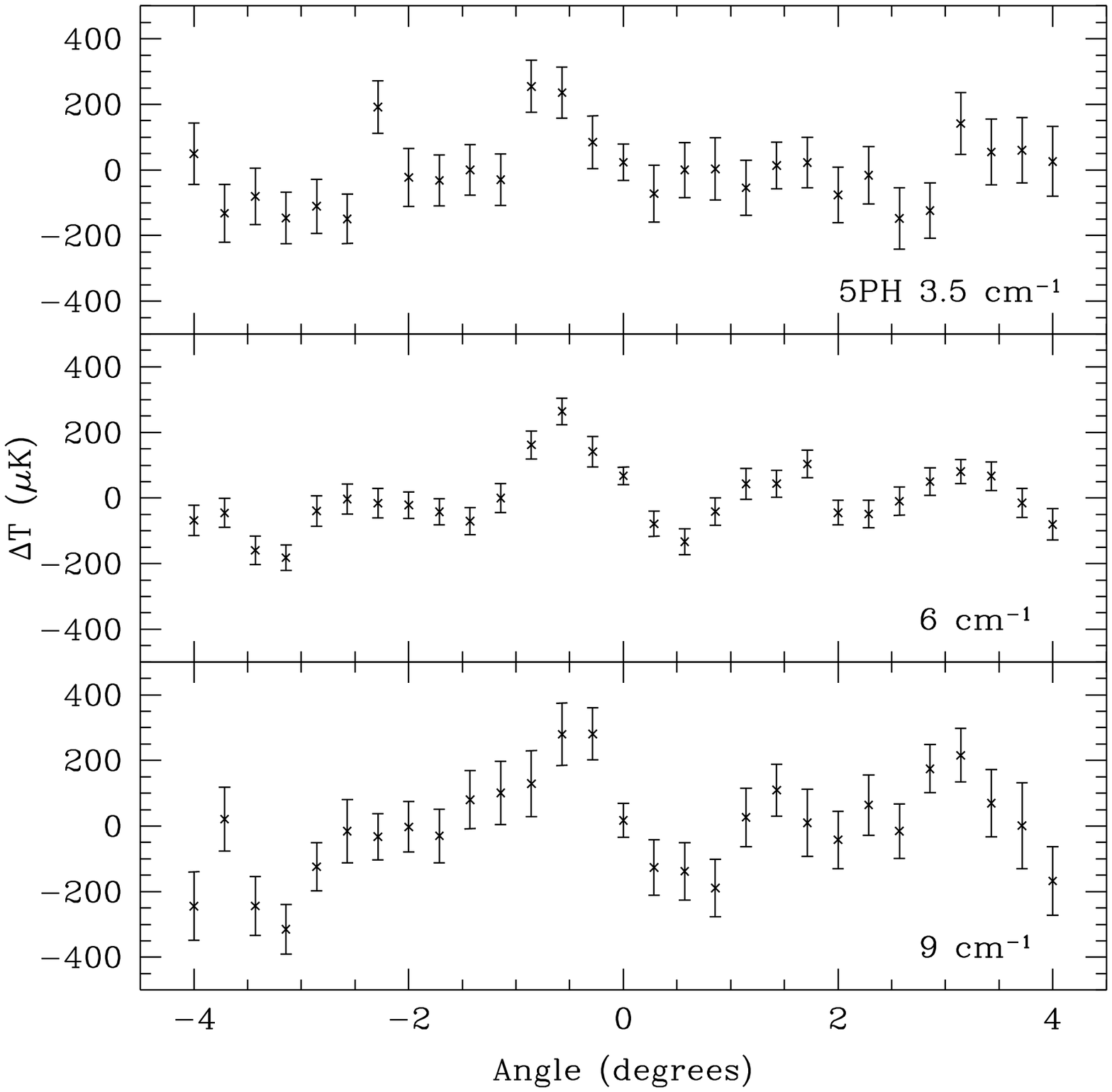}
\end{center}
\vfill
Figure~\ref{fig:data}e

\clearpage

\begin{center}
  \leavevmode
  \epsfxsize=6.5truein
  \epsfbox{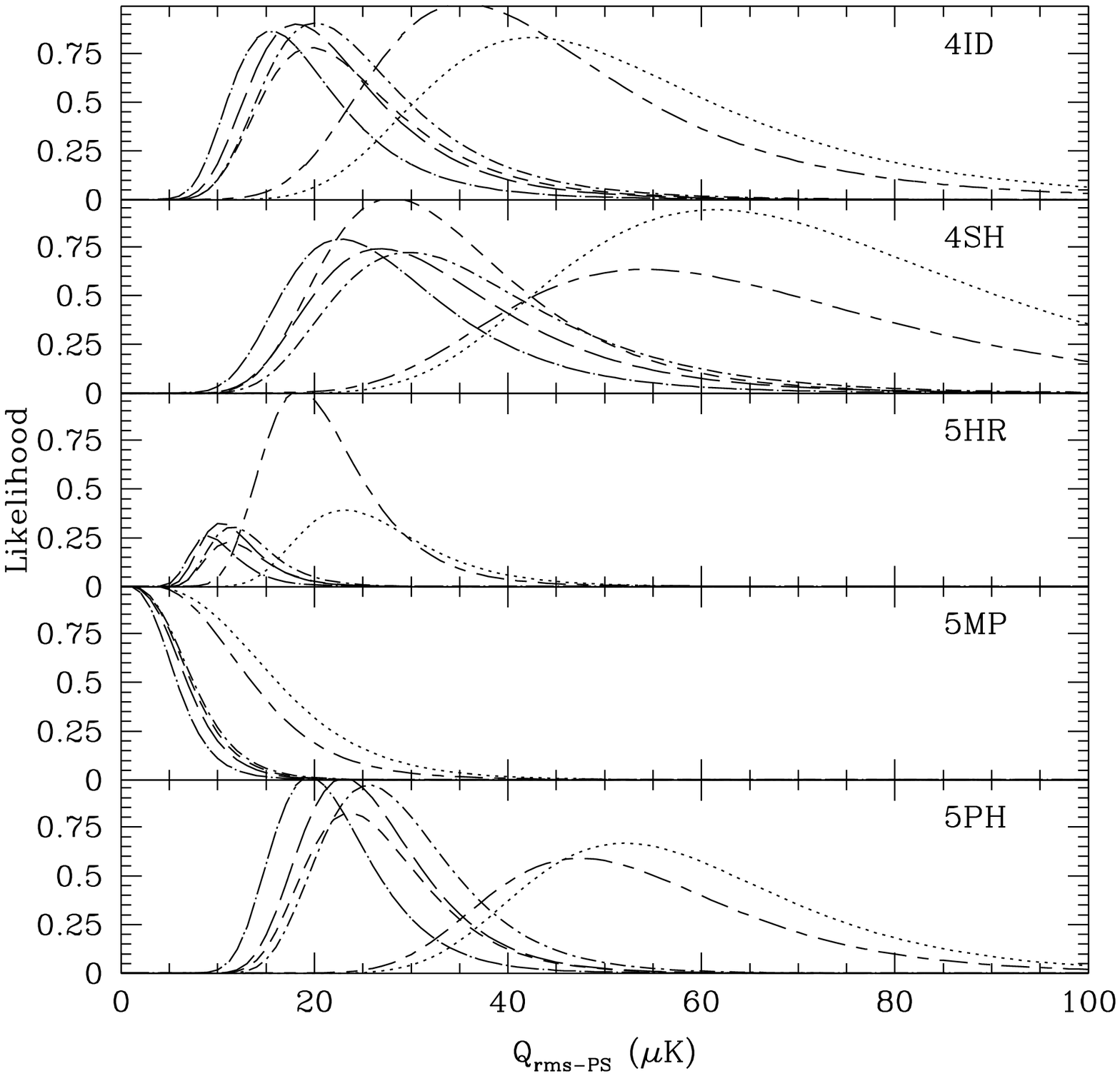}
\end{center}
\vfill
Figure~\ref{fig:likenoms}a

\clearpage

\begin{center}
  \leavevmode
  \epsfxsize=6.5truein
  \epsfbox{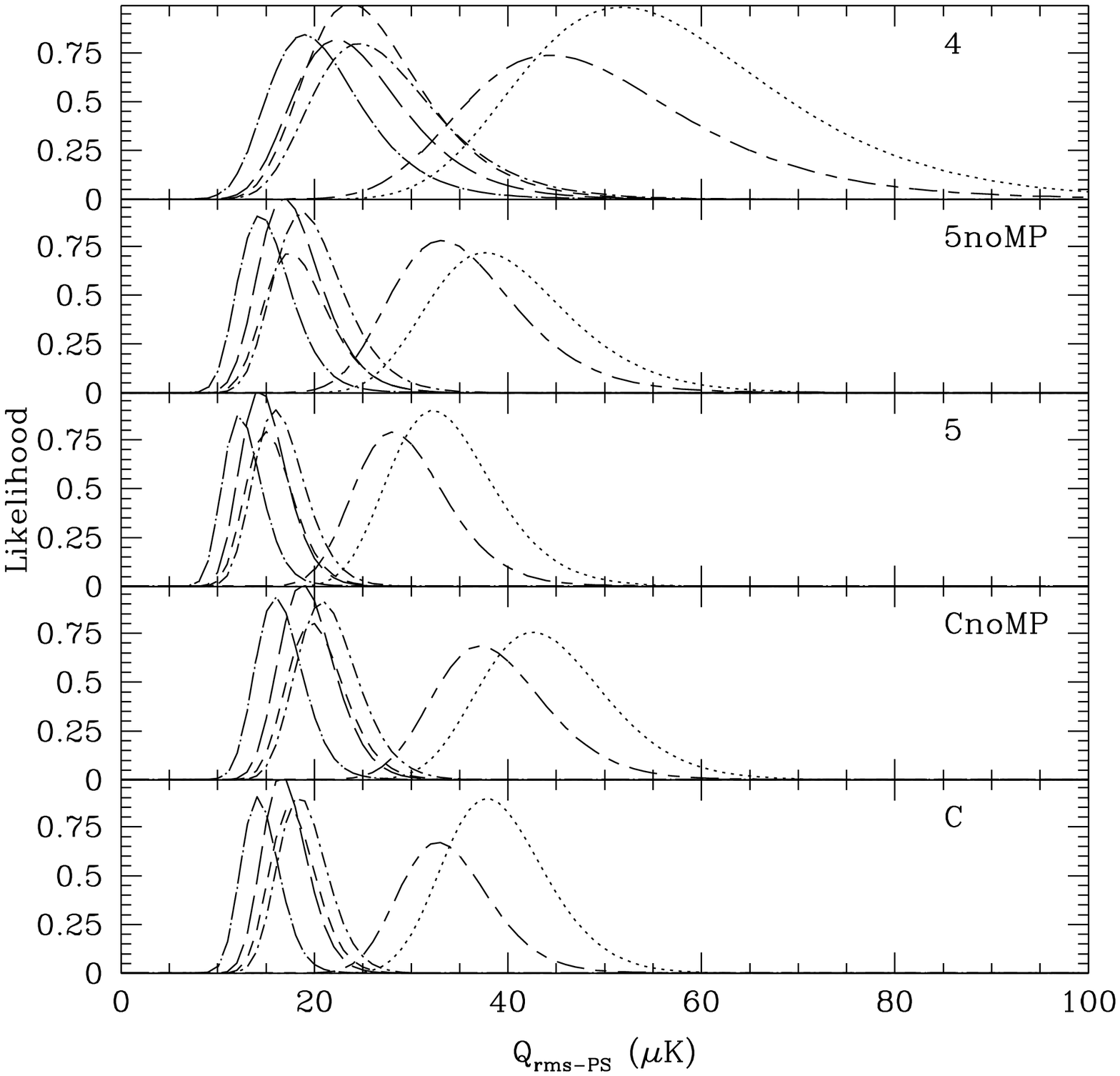}
\end{center}
\vfill
Figure~\ref{fig:likenoms}b

\end{document}